\documentclass[letterpaper]{article} % DO NOT CHANGE THIS
\usepackage{aaai24}  % DO NOT CHANGE THIS
\usepackage{times}  % DO NOT CHANGE THIS
\usepackage{helvet}  % DO NOT CHANGE THIS
\usepackage{courier}  % DO NOT CHANGE THIS
\usepackage[hyphens]{url}  % DO NOT CHANGE THIS
\usepackage{graphicx} % DO NOT CHANGE THIS
\usepackage{natbib}  % DO NOT CHANGE THIS AND DO NOT ADD ANY OPTIONS TO IT
\usepackage{caption} % DO NOT CHANGE THIS AND DO NOT ADD ANY OPTIONS TO IT
\usepackage{algorithm}
\usepackage{algorithmic}

\usepackage{newfloat}
\usepackage{listings}
\DeclareCaptionStyle{ruled}{labelfont=normalfont,labelsep=colon,strut=off} % DO NOT CHANGE THIS
\floatstyle{ruled}
\newfloat{listing}{tb}{lst}{}
\floatname{listing}{Listing}
\title{Foregrounding Artist Opinions: A Survey Study on Transparency, Ownership, and Fairness in AI Generative Art}
\author {
     Juniper Lovato \textsuperscript{\rm 1,2}, Julia Witte Zimmerman \textsuperscript{\rm 2}, Isabelle Smith \textsuperscript{\rm 3}, Peter Dodds \textsuperscript{\rm 1,2}, Jennifer L. Karson \textsuperscript{\rm 4}\\
}
\affiliations {
    \textsuperscript{\rm 1}Department of Computer Science, University of Vermont\\
    \textsuperscript{\rm 2}Vermont Complex Systems Center, University of Vermont\\
    \textsuperscript{\rm 3}Department of Mathematics \& Statistics, University of Vermont\\
    \textsuperscript{\rm 4}Department of Art and Art History, University of Vermont \\ 
    jlovato@uvm.edu
}
\usepackage{bibentry}
\begin{document}

\maketitle

\begin{abstract}
Generative AI tools are used to create art-like outputs and sometimes aid in the creative process. These tools have potential benefits for artists, but they also have the potential to harm the art workforce and infringe upon artistic and intellectual property rights. Without explicit consent from artists, Generative AI creators scrape artists' digital work to train Generative AI models and produce art-like outputs at scale. These outputs are now being used to compete with human artists in the marketplace as well as being used by some artists in their generative processes to create art. We surveyed 459 artists to investigate the tension between artists' opinions on Generative AI art's potential utility and harm. This study surveys artists' opinions on the utility and threat of Generative AI art models, fair practices in the disclosure of artistic works in AI art training models, ownership and rights of AI art derivatives, and fair compensation. Results show that a majority of artists believe creators should disclose what art is being used in AI training, that AI outputs should not belong to model creators, and express concerns about AI's impact on the art workforce and who profits from their art. We hope the results of this work will further meaningful collaboration and alignment between the art community and Generative AI researchers and developers.  
\end{abstract}

\section{Introduction}
\label{section:intro}

Art has a rich history in co-evolution with the development of new technologies~\cite{etro2023history}. The advent of Artificial Intelligence (AI) tools used to create or assist in the creative process carries great potential utility for artists but also harbors the significant potential for harm to the art workforce and artistic and intellectual property. To examine this tension between utility and threat, we surveyed 459 artists about their views on AI-generated art. The survey results reveal the nuanced perspectives of artists on Generative AI. When the discussion is thorough and directly related to their art practice, it prevents their views from being reduced to a simplistic binary sentiment, which fails to represent the diversity of their field. By capturing artists' views through a survey, this method moves beyond anecdotal evidence and provides a counterbalance to third-party opinions as well as those of the Generative AI industry.

% uncomment out for arxiv 
\begin{figure}[t]
     \centering
         \includegraphics[width=0.85\columnwidth]{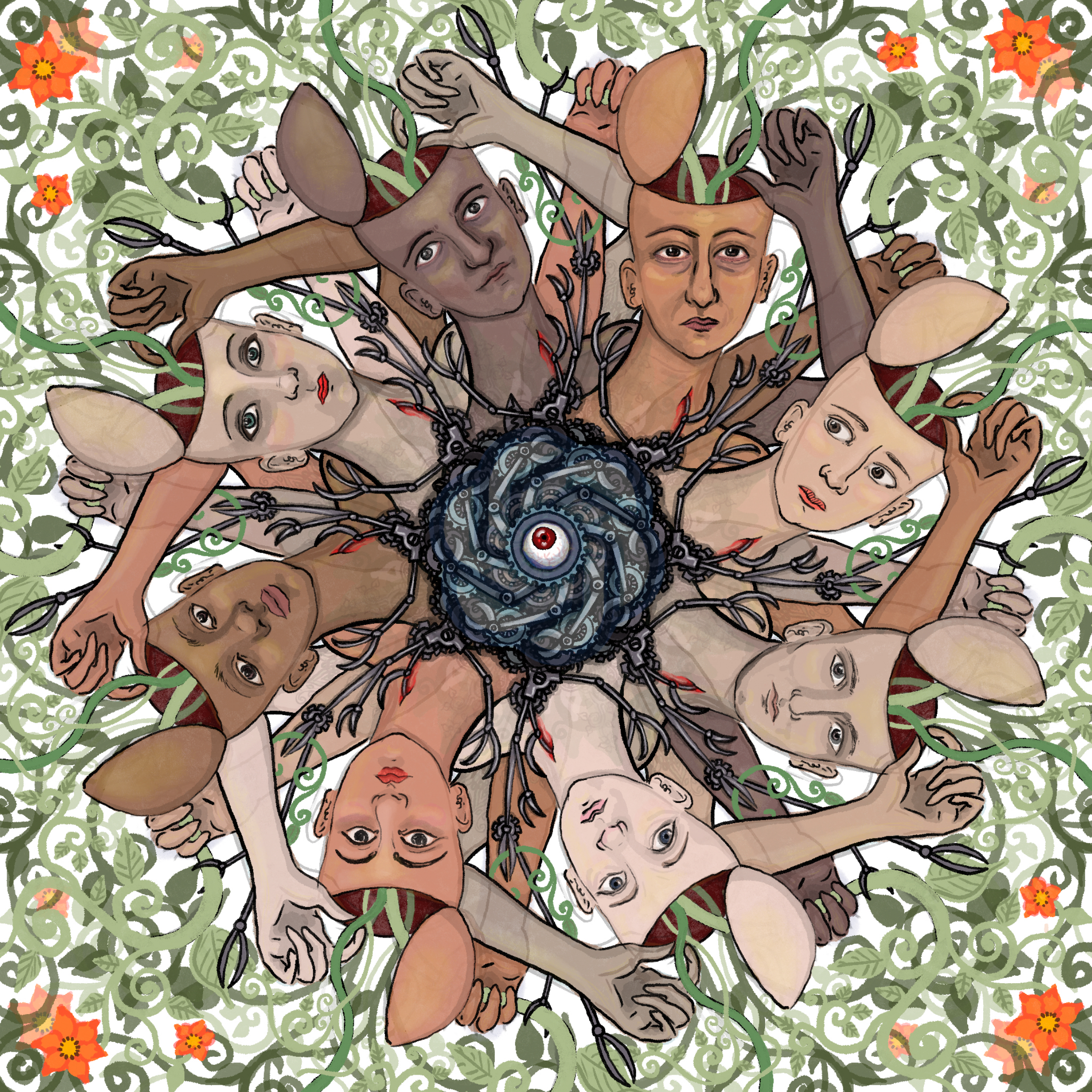}
        \caption{``Art is the joint responsibility of all [people]'' ~\cite{larousse1970ancient}.}
        \label{fig:teaser}
\end{figure}

Our study surveys artists' opinions on fair practices in the disclosure of artistic works in AI training models, ownership and rights of AI art derivatives, and fair compensation models. Future work will involve conducting long-form interviews with them to understand artists lived experiences. Our research questions for this study can be summarized as follows: RQ1: Do artists see AI models as a threat to art workers?; RQ2: Do artists see AI art models as a positive development in the field of art?; RQ3: What types of disclosure for the use of artwork to train an AI art model do artists consider fair?; RQ4: From an artist's perspective, who should own the derivative artwork created by AI art models? 

We summarize our research results as follows: (Q1) 61.87\% of survey participants agreed that AI models are a threat to art workers. (Q2) 44.88\% of survey participants agreed that AI art models are a positive development in the field of art (22.03\% of survey participants agreed that both AI was a positive development and a threat). (Q3) Artists overwhelmingly agreed (80.17\%) that model creators should be required to disclose in detail what art and images they use to train their AI models. (Q4) Artists were split between thinking that derivatives from AI art models should be owned by the original artists whose work is represented in the AI art (41.39\%) and the user who used the model to create the AI art (39.22\%). In our study, 26.80\% of the participants agreed that the work and its derivatives should be considered the property of the AI model creators. We also looked into what types of compensation artists consider to be fair in exchange for the use of their artwork to train an AI art model. Half of the respondents (50.97\%) stated that they did not need compensation, but what mattered to them was who was profiting from the use of their artwork. The most common fair compensation option selected by the survey participants was ``I don't need to profit, but I don't want for-profit companies to profit from my art'' (22.80\%). This suggests that artists are not entirely against their art being used in AI art model training data but oppose for-profit companies benefiting financially. 

%We describe the survey results in detail in the Results Section and determine which characteristic variables of our survey participants are most associated with these results. We will present the survey methodology in the Methods Section. 

Our paper draws on our team's multidisciplinary background -- in art and art history, social science, complex systems, computer science, mathematics, and statistics -- to address these questions and understand the possible impact of Generative AI on artists. We hope this work will provide a modest step towards understanding artists' opinions on Generative AI technology, its impact on their field, and the role they would like their work to play in the burgeoning field of Generative AI. 

\section{Background}
\label{section:background}

There is ongoing disagreement regarding whether the output of Generative AI can be considered art~\cite{ftcroundtable2023, jiang2023aiart}; we will not examine the topic in depth here. For convenience, we'll refer to AI-generated output as ``art'' in this paper without asserting or refuting its artistic merit.

Jiang et al. define art as a human endeavor linked to culture and experience~\cite{jiang2023aiart}. They argue that human creativity, distinguished by unique inspiration and corporeal engagement, differs from AI-generated work. Recent court decisions in the US have similarly denied copyright protection to autonomously produced AI creations, underscoring the role of human creativity in copyright law~\cite{howell2023civil}.

Three compelling arguments against considering AI output as art emerge: art's human origin devoid of "organic pressure," the intimate connection between artist and world, and art's unique role in human experience~\cite{jiang2023aiart, larousse1970ancient, pb_bodyshapes}. Regardless of the artistry debate, AI-generated creations can meet commercial standards, like book covers, entering diverse artistic spheres, and impacting practitioners.

Artistic use of technology and AI tools~\cite{10.1145/3475799} is not novel; artists historically adopt tools, including machines and automation, often innovating beyond standard applications~\cite{larousse1970ancient}. Artists engaging with technology question its role in art and explore its boundaries, raising queries about intentionality, distraction, and artistic agency~\cite{karson2023guardrails}.

It is important to note that, although how technology should be incorporated into art may be up for debate, that technology is part of the artistic process is not new or ambiguous \cite{mazzone2019art}. Artists have always used tools~\cite{larousse1970ancient}, and many art forms involve machines (such as woodwork, metalwork, and digital art) and automation (such as digital printers and cameras). Some artists also choose to use technology in ways other than its intended purpose or develop new art-making techniques by breaking technology. Such artists may be both suspicious and intrigued by the technology they use, questioning the relationship between humans and machines and expanding art practices and processes. When presented with new technology, an artist may ask questions like ``How can I keep my work with the machine intentional? How might this technology distract me? Is the machine using me, or am I using it? Is this image-making or art-making? How does this automation serve my artistic intentions and practice?''~\cite{karson2023guardrails}.

In the case of machine learning, a crucial question for many artists is ``What data was used to train this model?'' Such inquiries are part of the artistic tradition of challenging the status quo and are in stark contrast to how the creators of Generative AI models have presented their tools for use to the general public.

\subsection{Generative AI Models}
\label{section:whatisagenerativeartmodel}

Generative AI models can generate outputs across various media types based on user inputs, such as text prompts or multimedia~\cite{gozalobrizuela2023survey}. Key players in textual Generative AI include large language models (LLMs) like OpenAI's ChatGPT and GPT-4, while visual models like StableDiffusion and Midjourney modify images using iterative noise adjustments guided by diffusion processes.

The field of Generative AI has sparked significant excitement due to its broad range of applications, including both anticipated ones valued in the tens of billions of dollars and substantial academic and industrial research efforts~\cite{bubeck2023sparks, abdin2023kitab, jiang2023aiart, stanfordaii2023index}. This expansive field encompasses thousands of models and tools that are driving research across various academic disciplines. However, alongside this enthusiasm, there is a pressing need for a more critical examination of the ethical implications associated with these tools and methods \cite{10.1145/3600211.3604686}. The deployment of these tools beyond academia has resulted in adverse consequences such as personal and financial losses, widespread information manipulation, and disruptions to sociotechnical infrastructure \cite{ferrara2024genai, 10.1145/3600211.3604715}.

For the purposes of our paper, we define Generative AI as being a subset of tools or processes that combine massive probability distributions with sense-making algorithms in order to create output that, taken as a whole, was not included in the training data~\cite{jiang2023aiart}. The probability distribution should be based on the training data. The sense-making algorithm could be based on several strategies, such as a physical process like diffusion or an architectural approach like transformers/neural nets, although in the end, it should produce something akin to a latent/embedding space; a place to hold meaning~\cite{gozalobrizuela2023survey, zhao2023survey, grand2022semantic}. Research is ongoing into the various distributions and processes that can be implemented in Generative AI models to achieve specific behaviors and results~\cite{xu2023pfgm}. However, irrespective of details, all of these models use advanced architectures in conjunction with large amounts of data gathered through brute-force methods across vast swathes of the publicly accessible internet~\cite{xu2023pfgm, zhao2023survey}.

In this paper, we are interested in how artists perceive Generative AI, what its impact on art-related careers might be, and what moral, ethical, artistic, and financial implications are coming hand-in-hand with the broader flourishing of Generative AI~\cite{stanfordaii2023index, gozalobrizuela2023survey, zhao2023survey}. Therefore, we touch on three relevant topics -- first, related academic work; second, the economic context for this discussion; and third, what sort of stages are within the pipeline of Generative AI models and how those stages can each cause harm (a specific and noteworthy type of impact) -- within this Background section.

Although in our paper we focus more on the pressing potential for harm, it is worth noting that there is also growing excitement around Generative AI's capabilities and how it could be incorporated into artistic processes~\cite{karson2023guardrails}. For example, artist Cosmo Wenman worked with blind researchers Brandon Biggs, Lindsay Yazzolino, and Joshua Miele to create images using the Midjourney tool. Their team noted that programs like Midjourney can find ``weird and thematic intersections'' in art that go unnoticed by human eyes. They also discussed the role of chaos in creating ``visual intrigue'' and how technology can be used to ``enhance agency.'' Wenman called the new technology ``essentially moveable type for visual rhetoric''~\cite{schrader2023blind}.

\subsection{Related Work}
\label{section:relatedwork}

A paper by Jiang et al. \cite{jiang2023aiart} analyzes Generative AI art models and their potential impact on the world through the lens of diverse academic sources (both philosophical and scientific). They describe many opportunities for harm as a result of such products: violation of the consent and privacy of artists~\cite{ho2024midjourney}, a lack of dataset transparency (shielding violations and biases)~\cite{ho2024midjourney}, a failure to uphold the contextual integrity of content~\cite{nissenbaum2004privacy}, harm to the livelihood of the artist, biases in the training data, automating processes in a way that is detrimental to the creation of art, reduced accessibility, erasure of human works by obscurity, and increasing reluctance by artists to share their works openly. 

The harms Jiang et al.~\cite{jiang2023aiart} foresee for artists fall into two categories, financial and emotional, which manifest as the replacement of a human artist on a project, the impersonation of an artist without consent~\cite{gozalobrizuela2023survey}, or plagiarism. Impersonation, in this context, could be weaponized for nefarious purposes~\cite{ho2024midjourney}. Moreover, these two kinds of harm, enacted against individuals at scale, create societal-level harm. Specific societal harms Jiang et al. discuss in their work include data laundering -- through academic or open-source projects where public funds are funneled into research with private profits -- the proliferation of stereotypes and ``hegemonic views,'' and erasure by obscurity (i.e., that AI-generated outputs can be created much more quickly than human ones). They also point out that US copyright and intellectual property jurisprudence needs to be better equipped to deal with these new tools and that open sourcing, often called on as a solution to software-related problems, is incomplete in this case~\cite{jiang2023aiart}. 

As possible solutions, Jiang et al. bring up regulations regarding opt-in consent (for training data), independent funding (much current research is supported by industry), and education about how power interacts with technology. In particular, they note there are many incentives to label work as ``apolitical'' and that the spin capitalism is putting on Generative AI is the ``democratization of art'' (and other fields)~\cite{jiang2023aiart}.

Going beyond Jiang et al.'s list of solutions, Shan et al. ~\cite{shan2023promptspecific} propose tools allowing artists to ``poison'' models that scrape their work without permission. The tool they created (Nightshade) can ``destabilize general features'' in the targeted generative model and ``bleed-through'' to semantically similar prompts without injecting large amounts of tailored data~\cite{shan2023promptspecific}.

\subsection{Economic Context: Livelihood of the Artist}
\label{section:economic}

AI's integration into the global economy is rapidly expanding, as evidenced by the increasing number of AI-related job postings across various economic sectors~\cite{stanfordaii2023index, squicciarini2021demand}. Employers worldwide are seeking AI-skilled workers, with many companies opting to adopt AI tools each year. The primary benefit reported by these companies is a reduction in costs, often achieved through layoffs and the substitution of human roles with AI-driven solutions~\cite{stanfordaii2023index, gmyrek2023generative, alekseeva2021demand}.

Some of these impacts are already coming to bear on the economic reality of artists \cite{chow2020ghost, henderson2010s}. Entry-level positions in film, TV, and gaming seem particularly at risk, and the conceptual period of a project seems to be a popular phase in which to replace artists with AI. Traditional niches like illustrating book covers can now be filled by the output of Generative AI, and major employers can reduce the number of employees needed (for example, Netflix Japan's layoff of animators)~\cite{jiang2023aiart}.

The economic impact on artists is not solely determined by the output of AI models. Harm, both financial and emotional\cite{caporusso2023generative}, can arise from the use of content without explicit consent. This includes situations where individuals may not have anticipated their work being repurposed by machine learning models or the entities controlling and utilizing these technologies~\cite{nissenbaum2004privacy}. For instance, consider an artist with a physical gallery presence whose work is photographed by a visitor and subsequently used in ways they did not foresee, especially in the intricate landscape of social media~\cite{ho2024midjourney}.

Moreover, the issue of copyright protection becomes paramount when AI models are trained using human artworks without proper consent, attribution, or adequate compensation for the original creators. This concern is shared among artists, creative industries, and government agencies alike~\cite{ftcroundtable2023, marcus2024minefield}.

\subsection{Problems in the Pipeline of Generative AI Tools}
\label{sec:pipeline}

As discussed earlier, when assessing the impact of an AI model, it extends beyond the usage of the model's output or its deployment (such as replacing human workers). Every stage involved in developing a Generative AI model can influence the eventual harm caused by the model, thereby affecting its overall impact. Although many stages in this pipeline are typically inaccessible, except for the input and output phases, they still play a significant role. In this section, we outline the pipeline of creating a Generative AI model and focus on one prevalent issue, bias~\cite{zhao2023survey}, to illustrate how various stages within the pipeline can contribute to the model's impact.

When developing an AI model, several steps are involved. Firstly, training materials need to be collected, often comprising a subset of real-world artifacts. These materials then undergo an assessment to address issues like poor quality, irrelevance, or over-representation of certain artifacts ~\cite{luccioni2021box, zhao2023survey}. Next, the model's architecture is assembled, followed by the training process. For complex Generative AI models with unpredictable behavior, there is an additional step called alignment. This step aims to constrain the model's range of potential behaviors to align with the creators' desired outcomes~\cite{zhao2023survey}. 

Early in the pipeline, the selection of training material significantly shapes the eventual behavior of the model. For instance, many important datasets used in machine learning benchmarks predominantly feature lighter-skinned subjects~\cite{buolamwini2018gendershades}, which indirectly introduces bias into the model. This type of global bias cannot be detected by locally examining each piece of training data for harmful content. Furthermore, datasets often contain objectionable and biased content, including racist hate speech and sexually inappropriate images~\cite{luccioni2021box}, even after undergoing cleaning procedures.

In the middle of the pipeline, within the internal workings of the trained model, even fundamental and abstract elements like the meanings of common English words have been discovered to encode gender biases~\cite{bolukbasi2016homemaker}. As we progress toward the end of the pipeline, the output generated by the model can exhibit biases as well.  For example, the Stable Diffusion model disproportionately generates images of white male executives, doctors, lawyers, judges, and law enforcement, and also disproportionately generates images of black and brown criminals and service workers~\cite{ghosh2023person}. Additionally, users have reported that Generative AI tools often generate images where they appear younger, more sexualized, or with lighter skin tones, even when these attributes were not requested or suggested in the input prompt ~\cite{ghosh2023person}. These examples are not unique; similar biases are found across current AI models and products -- including those of Generative AI -- at every level, from the training data they learned from to their internal representations to their outputs~\cite{luccioni2021box, ghosh2023person, cheng2023marked, zhao2023survey}.

As a final problem in the pipeline, it is worth noting that these AI tools cannot be created ex nihilo: they are integrated into complex human systems that bring their own forces to bear on the model's potential impact. There are discrepancies in the demographics of who is wielding the power of these new technologies: who is making them~\cite{stanfordaii2023index}, profiting off them, using them. This can have many knock-on effects, sometimes through mechanisms difficult to foresee~\cite{even-tov2023sharing}, with respect to the design, use, and role of these tools in society~\cite{jiang2023aiart, stanfordaii2023index}. In particular, the companies behind these products often employ a great deal of secrecy as to their training data, architecture, or implementation (or all three)~\cite{ho2024midjourney}. As a result, users must interact with the product as a black box~\cite{jiang2023aiart, marcus2024minefield}; concerned users can only conjecture as to what kind of training data could produce the observed results based on what is known of the basic architecture of the model, which in turn allows argument over the obscure details. The lack of transparency benefits the companies offering these products (with respect to establishing the impact, specifically the harms, of the model) in giving them a way to sidestep moral and ethical concerns~\cite{marcus2024minefield}.

\section{Methods}

\subsection{Survey Methodology}
\label{section:survey}

Our anonymous original survey of artists is an observational study (initial N=516 (includes incomplete answers), final N=459 (only complete answers)) of artists over the age of 18 years that was collected using a convenience sample in phase 1 and supplemented with a Qualtrics survey panel in phase 2 (demographic breakdown can be seen in Table \ref{tab:surveydescstats1}. Qualtrics is a reputable academic survey panel platform that has been shown to yield reliable results in comparison to other survey platforms \cite{boas2020recruiting}. A comparison of the distributions of the US art workforce) based on the demographics is presented in Appendix B. Survey participants were recruited based on their self-identification as artists and must be over the age of 18. All survey questions are original and developed by the authors.  Incomplete surveys were dropped from the analysis, leaving a final N of 459. Participants provided consent by agreeing to the survey information sheet. They were then prompted to respond to inquiries regarding their artistic identity and financial interactions within the art domain. Participants were presented with a brief introduction to AI-generated art, including an illustrative example, and were asked to indicate their familiarity with AI art tools and models. Subsequently, participants were queried on their perceptions regarding ownership of art produced by AI systems using a 5-point Likert scale. Further questions focused on artists' preferred compensation criteria for their work being used in AI-generated art models. Participants were also asked to evaluate the impact of AI art models on their field, their perception of these models as potential threats, and whether they believed creators of such models should disclose their data sources using a 5-point Likert scale. Finally, demographic information about the participants' backgrounds was collected. An appendix that enumerates all survey questions is available in Appendix A, which can be found in our supplementary materials \cite{lovato2024foregrounding}.  This survey was approved as ``exempt'' by the University of Vermont Institutional Review Board (IRB).

Sampling limitations: In future work, we would like to sample a much bigger representative sample that specifically targets artistic practices and demographics. In addition, we are interested in how the dynamics of these opinions change over time with surveys that assess artist opinions at regular intervals. 

\textbf{Phase 1 Sampling}: Survey participants for phase 1 (n = 252) of our observational study were recruited broadly on social media, using flyers locally at the University of Vermont, emailing artist groups, and direct contacts who are professional artists. Survey participants who are over the age of 18 and identify as an artist or work in an arts-related field were eligible. Phase 1 of the surveys was collected through a convenience sample from March 25, 2023, through September 8, 2023. Survey responses were collected on Qualtrics, an online survey platform. 

\textbf{Phase 2 Sampling}: Phase 2 (n = 264) was collected via a targeted survey panel conducted by Qualtrics from September 12-23, 2023 on the Qualtrics online survey platform. The inclusion criteria for the study in Phase 2 were the same as in Phase 1. 

\subsection{Data}
\label{section:data}

Our original survey of artists is an observational survey that took place over two sample phases (total N=459 completed surveys over both phases). We transformed all survey response variables of interest into numerical form to analyze our survey results (survey questions 8,9,10,13,14,15 as seen in Appendix A). All Likert survey questions were converted from `Strongly Agree,' `Agree,' `Neutral,' `Disagree,' and `Strongly Disagree' to an ordinal scale of 1,2,3,4,5. For our ordinal logistic regression analysis, we combine  `Strongly Agree' and `Agree' to be equal to 3, 'Neutral' to be equal to 2, and `Disagree' and `Strongly Disagree' to be equal to 1. In the original logistic regression, we reduced the dimensionality of the Likert scale to investigate the positive, neutral, and negative sentiment of the response questions in relation to our secondary variables. 

All nominal and categorical variables (survey questions 1,2,3,4,5,11,12,16,17,18,19, as seen in Appendix A) were transformed into binary variables. Survey questions that included an ``other'' option or write-in answers were also categorized (e.g., in question 2, we added an additional art practice, which included digital art from the write-in answers). 

Demographic questions (survey questions 16, 17, 18, 19 as seen in Appendix A) included asking survey participants their gender identity to identify whether they were white or a person of color, choosing their age range and country of residence. All demographic questions were transformed into binary variables. 

\subsection{Analytical Methods}
\label{section:methods}

\textbf{Odds ratio}
In general, the odds of an event are the probability of success over the probability of failure. This can be written as the following equation: \begin{equation} \frac{p}{1 - p} \end{equation} where p is the probability of success. Unlike probability, odds can be from 1 to positive infinity ~\cite{bruin2011interpret}. The odds ratio (OR, or $\theta$) is simply the odds of one event divided by the odds of another. The OR can be from 0 to positive infinity, with an OR equal to 1 indicating equal odds of success between events. The natural logarithm of the OR is often used, as the transformation renders the distribution symmetric about 0 ~\cite{agresti2012categorical}.

\textbf{Ordinal Logistic Regression}
Ordinal logistic regression (OLR) is a generalization of regression (either multiple linear or binomial logistic) used for an ordered categorical response variable and any number of categorical or numeric secondary variables. OLR finds the logit (also known as log-odds) of the dependent variable (Y) being less than or equal to a certain level (denoted j)~\cite{bruin2011interpretorl}. which can be written as: \begin{equation}logit(P(Y \leq j)) = log\frac{P(Y \leq j)}{P(Y > j)}\end{equation} It is important to note that OLR assumes proportional odds, meaning the effect of the secondary variable(s) is the same between all levels of the dependent variable.

\section{Results}
\label{section:results}

\begin{table}[t]
\small
\centering
\tabcolsep=0.19cm
\begin{tabular}{@{}*{2}{l}@{}}
\textbf{Type} & \textbf{Count (\%)} \\ \hline
Male & 273 (59.48\%)\\
Female  & 163 (35.51\%) \\
Non-binary & 23 (5.01\%)\\
White & 361 (78.65\%) \\
Person of Color & 98 (21.35\%)\\
Age 30-49 & 233 (50.76\%)\\
Age 18-29 & 104 (22.66\%)\\
Age 50-64 & 79 (17.21\%)\\
Age 65 or older & 43 (9.37\%)\\
\hline
\end{tabular}
\caption{Descriptive statistics for our survey participants, N=459. A detailed table of descriptive statistics can be seen in  Appendix B in Table \ref{tab:surveydescstats2}}
\label{tab:surveydescstats1}
\end{table}

\begin{table*}[t]
\small
\centering
\tabcolsep=0.19cm
\begin{tabular}{@{}*{6}{l}@{}}
\textbf{Question} & \textbf{Strongly Disagree} & \textbf{Disagree} & \textbf{Neutral} & \textbf{Agree} & \textbf{Strongly Agree} \\ \hline
AI threat to art workers & 6.32\% &	14.16\%	& 17.65\% &	33.77\%	& 28.10\%\\
AI positive development  for art & 13.07\%	& 15.03\% &	27.02\%	& 22.88\% &	22.00\%\\
Required disclosure	& 2.83\% &5.23\% & 11.98\% & 32.03\% & 48.15\% \\
Output owned by model creators & 32.46\% &	24.62\% & 18.74\% &	15.25\% & 11.55\%\\
Output owned by AI user & 19.61\% &	 22.22\% & 21.57\% & 20.92\% & 18.30\%\\
Output owned by the artist & 9.37\% & 26.58\% & 25.05\%	& 22.00\% &19.39\%\\
\hline
\end{tabular}
\caption{Participant answers to our primary research questions by percent (answered on a 5-point Likert scale), N=459.}
\label{tab:likertdescrstats}
\end{table*}

\begin{table*}
\small
\centering
\tabcolsep=0.19cm
\begin{tabular}{@{}*{2}{l}@{}}
\textbf{Type} & \textbf{ Counts (\%)} \\ \hline
I don’t need to profit, but I don't want for-profit-companies to profit from my art &	106 (22.8\%) \\
I would donate the use of my artwork to train AI art models (no compensation) &	82 (17.63\%)\\
A flat fee	& 55 (11.83\%)\\
A portion of any profit made from the model creators &	51 (10.97\%)\\
Not comfortable with any listed options  &	42 (9.03\%)\\
A portion of any profit from people who profit from derivative works made using the model &	32 (6.88\%)\\
A portion of any profit made &	32 (6.88\%)\\
I don’t need to profit, but I don’t want anyone else to profit from my art	& 27 (5.81\%)\\
I don't need to profit &	22 (4.73\%)\\
Other &	10 (2.15\%)\\
Tax & 6 (1.29\%)\\
\hline
\end{tabular}
\caption{Counts percent per compensation type.}
\label{tab:surveydescstatscomp}
\end{table*}
459 participants were anonymously surveyed, yielding a sample reflective of the demographics (gender, ethnicity, age) of the broader American adult art workforce (See table \ref{tab:surveydescstats1}). We conducted this observational study on the opinions of artists to explore the dynamic interplay between their perspectives on the potential benefits and risks of Generative AI art. This investigation delves into artists' viewpoints regarding the utility and risks associated with Generative AI art models, ethical considerations in disclosing artistic works used in AI art training models, the ownership and rights pertaining to AI art derivatives, and equitable compensation. Our results indicate a strong sentiment among artists in favor of disclosing what art is being used in AI training (RQ3), advocating against generative AI companies retaining ownership of AI-generated outputs (RQ4), and expressing concerns regarding AI's impact on the art industry's workforce (RQ1-2) and who profits from their art.

In the analysis of survey data pertaining to our primary research inquiries (see Table \ref{tab:likertdescrstats}), 80.17\% of the survey participants agree that model creators should be required to disclose in detail what art and images they use to train their AI models. We also find that 61.87\% of survey participants agree that AI models are a threat to art workers. In contrast, 44.88\% of survey participants agree that AI art models are a positive development in the field of art, notably with 27.02\% of artists selecting neutral on the Likert scale for this question. Interestingly, we see an overlap between artists who agree that AI art is a positive development and those who also agree that it is a threat to art workers; 22.03\% of survey participants agreed that both AI was a positive development and a threat, which presents an interesting tension which highlights that the sentiments of artists are nuanced and some hold complex opinions towards generative AI.

We examine a survey question prompting participants to consider a scenario where an AI art model was employed by another individual to create artwork distinctly resembling the participant's style and reflect on the rightful ownership of the resulting work and any derivative creations. Only 26.80\% of the participants agreed that the work and its derivatives be considered the property of the AI model creators, while 39.22\% of participants agreed that the work and its derivatives be considered the property of the person who used the AI model to generate the artwork. Finally, 41.39\% of the participants agreed that the work and its derivatives be considered the property of the artists whose style is being represented by the AI model output. This result presents an interesting tension as well between artists who think the ownership should belong to the original inspiration of the AI art while others may also want to retain the ownership of the resulting work. We see in our results that artists who have used AI art models are more likely to agree that the resulting work should be considered the property of the person who used the model to create the work. 

In the following sections, we will present results that identify the variables most strongly associated with survey participants who expressed agreement with the primary research questions.

We also look at what types of compensation artists consider fair in exchange for the use of their artwork to train an AI art model. The results for this question can be seen in Table \ref{tab:surveydescstatscomp}. We find that 50.97\% of survey respondents stated that they did not need compensation, but what matters to them is who is profiting from the use of their artwork. The fair compensation option selected most by our survey participants was ``I don't need to profit, but I don't want for-profit-companies to profit from my art'' (22.80\% selected this option), which indicates that artists are not wholesale against art being used in AI art models training data, but they oppose for-profit-companies reaping the financial benefits. Our results show that 36.56\% of respondents did state they would find some type of compensation fair, only 1.29\% of participants felt that a tax on companies and individuals who profit from the AI art outputs would be a fair option, and 11.18\% of participants felt that they would not feel comfortable with any of the compensation options listed. 

Descriptive statistics about our survey responses to demographic questions can be seen in Table \ref{tab:surveydescstats1}. A full table of descriptive statistics that outlines participants who have used AI art models, their familiarity with Generative AI, professional and artistic status, whether or not they have purchased art, and art practices can be seen in Table \ref{tab:surveydescstats2}.

\begin{figure*}
     \centering
         \includegraphics[width=0.85\columnwidth]{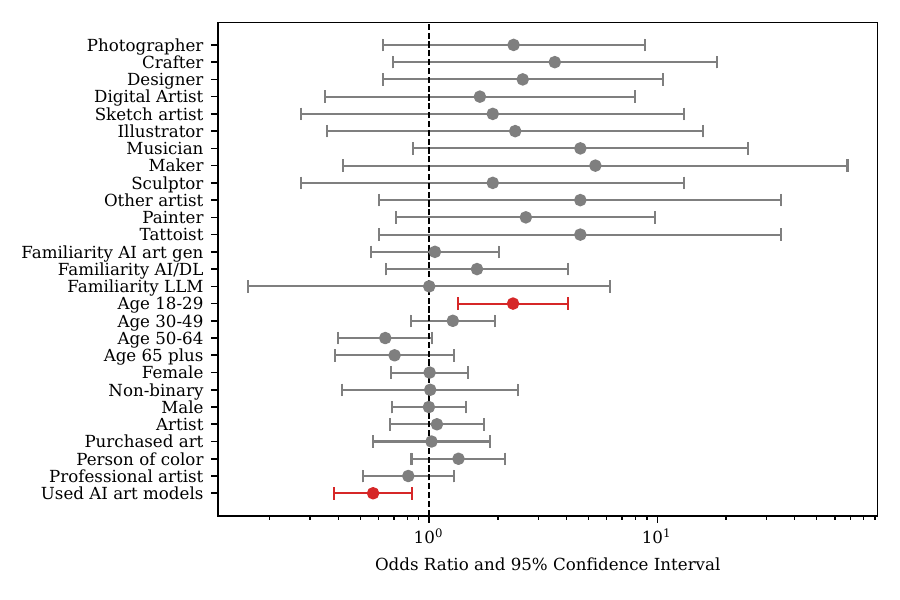}
         \includegraphics[width=0.85\columnwidth]{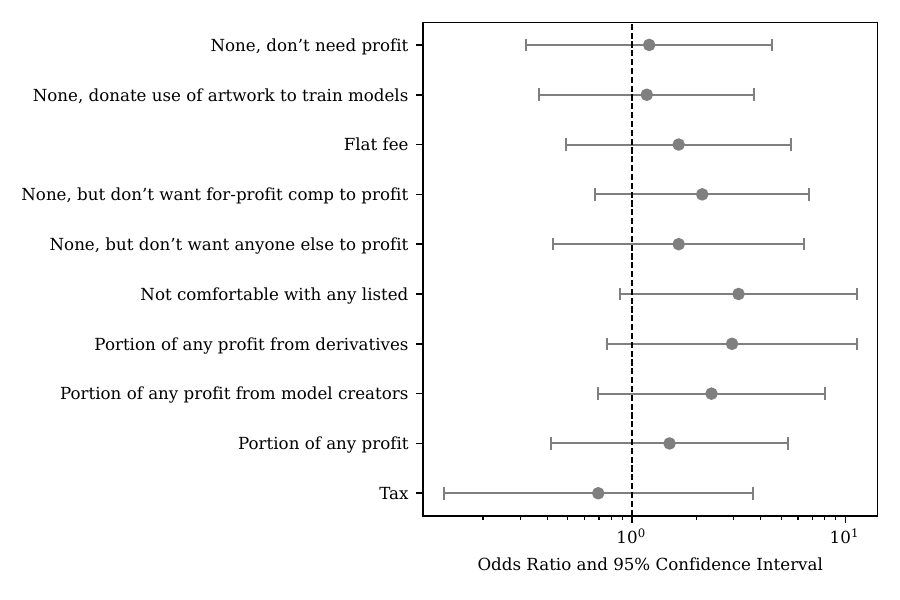}
        \caption{The ordinal logistic regression results show the association between answering `Agree' or `Strongly Agree' to RQ1 (Generative AI art models as a threat to art workers) and other survey variables. Results are represented by odds ratios on a log scale on the x-axis, with bars conveying a 95\% confidence interval and red indicating a p-value $<$ 0.05 and a confidence interval that does not cross the threshold of the dotted line in the figure.}
        \label{fig:RQ1fig}
\end{figure*}

\begin{figure}
     \centering
         \includegraphics[width=0.85\columnwidth]{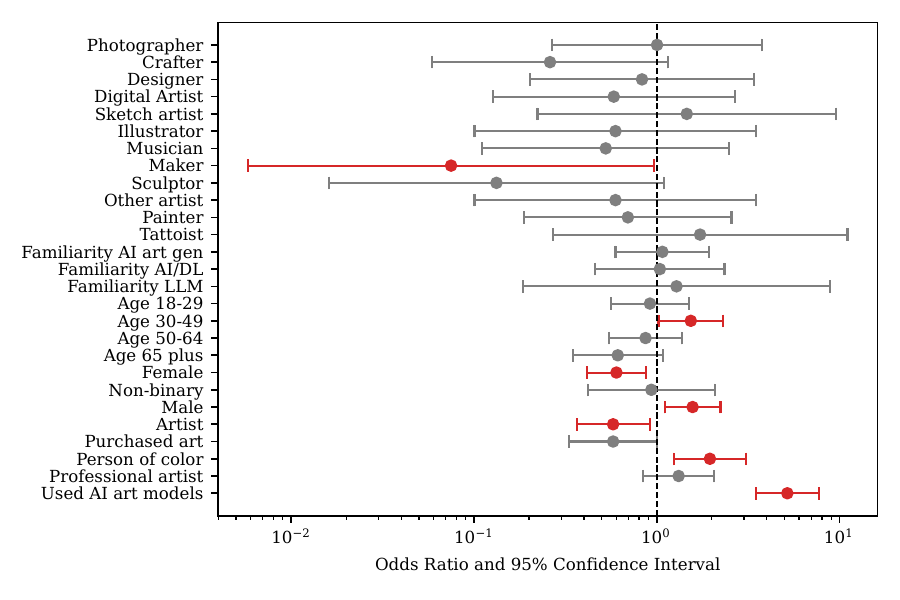}
         \includegraphics[width=0.85\columnwidth]{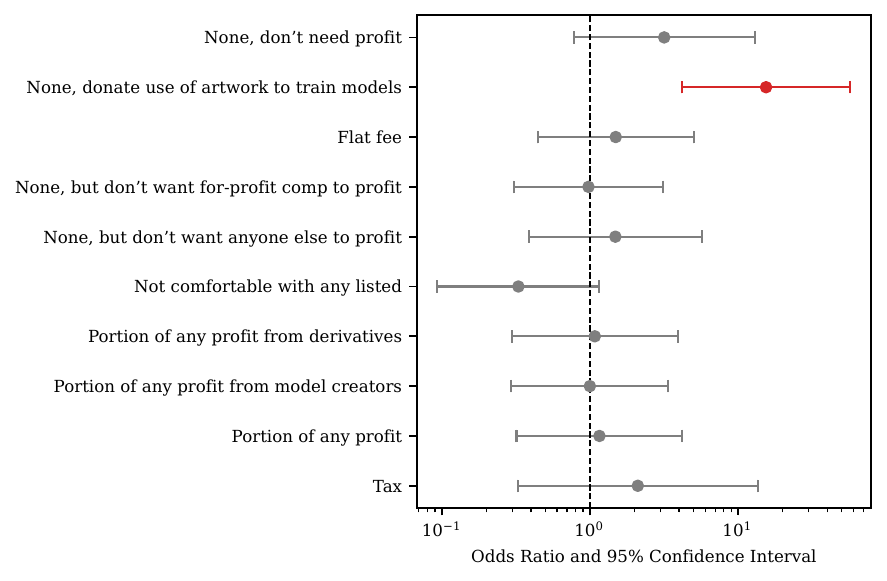}
        \caption{The ordinal logistic regression results show the association between answering `Agree' or `Strongly Agree' to RQ2 (Generative AI art models as a positive development) and survey secondary variables. Results are represented by odds ratios on a log scale on the x-axis, with bars conveying a 95\% confidence interval and red indicating a p-value $<$ 0.05 and a confidence interval that does not cross the threshold of the dotted line in the figure.}
        \label{fig:RQ2fig}
\end{figure}

\begin{figure}
     \centering
         \includegraphics[width=0.85\columnwidth]{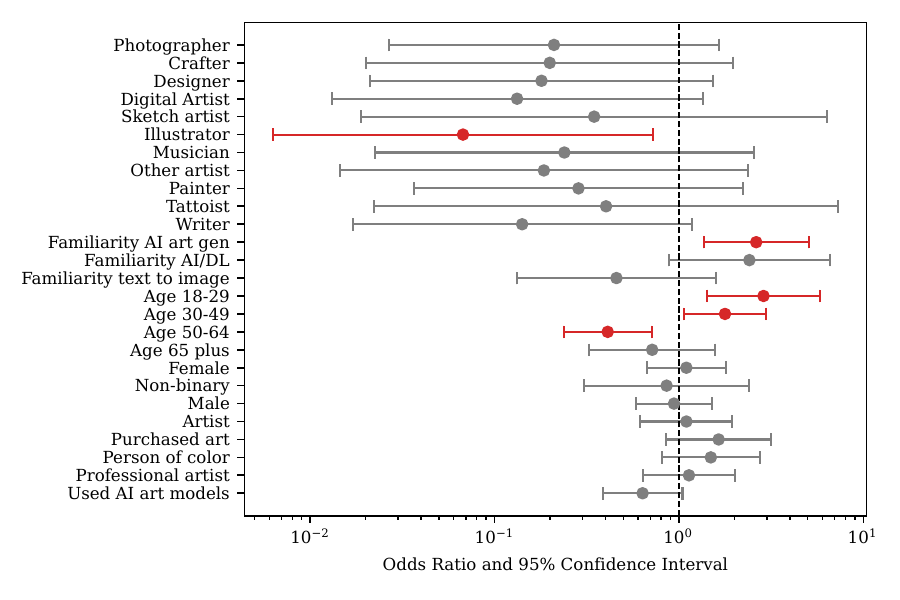}
         \includegraphics[width=0.85\columnwidth]{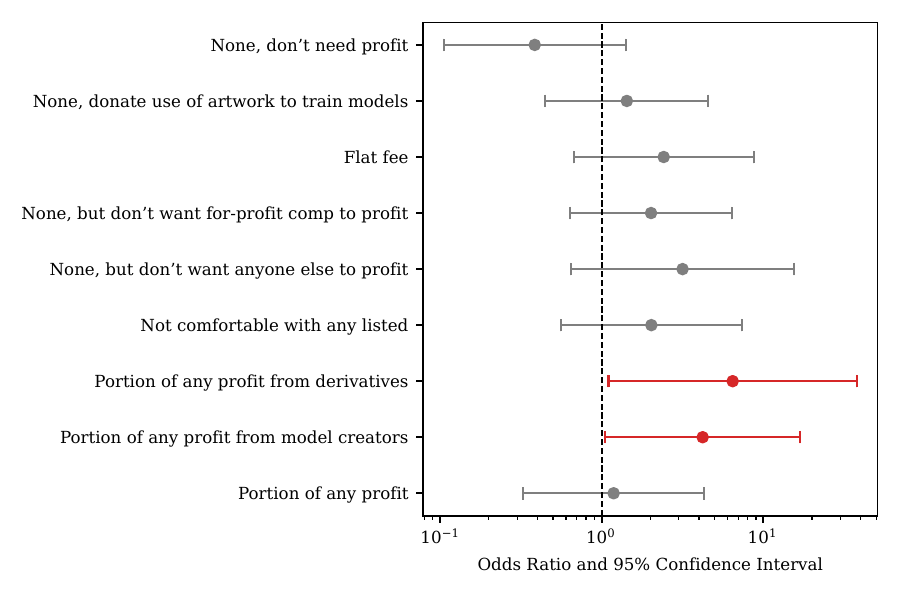}
        \caption{The ordinal logistic regression results show the association between answering `Agree' or `Strongly Agree' to RQ3 (Should model creators be required to disclose in detail what art \& images they use to train their AI models) and survey secondary variables. Results are represented by odds ratios on a log scale on the x-axis, with bars conveying a 95\% confidence interval and red indicating a p-value $<$ 0.05 and a confidence interval that does not cross the threshold of the dotted line in the figure.}
        \label{fig:RQ3fig}
\end{figure}

\textbf{RQ1: Do artists see Generative AI models as a threat to art workers?} 
\label{section:RQ1}

This research question investigates artists' perceptions of Generative AI models as potential threats to art workers. Our findings reveal that 61.87\% of survey participants hold the view that Generative AI models pose a threat to art workers. Thus, while a majority of artists perceive AI as a threat, it is not an overwhelming majority. This finding is related to survey question no. 15 (see Appendix A for survey questions). 

An ordinal logistic regression (OLR) analysis was performed to examine the relationship between the Likert scale responses to RQ1 (dependent variable) and the secondary variables (all odds ratios can be seen in Appendix C), focusing on the odds ratios derived from the OLR model (as seen in Figure \ref{fig:RQ1fig}). 

The results demonstrate a significant relationship (p-value  $<$ 0.05, all detailed odds ratios can be seen in Appendix C) between agreeing (represents a combination of both the responses `Agree' and `Strongly Agree' in the Likert scale) that Generative AI models are a threat to art workers and a subset of our secondary variables. Here, we see that artists ages 18-29 are more likely to agree that there is a threat and that artists who have used Generative AI art models to create art are less likely to think that it is a threat to the art workforce.

\textbf{RQ2: Do artists see Generative AI art models as a positive development in the field of art?} 
\label{section:RQ2}

This research question investigates artists' perceptions of AI models as positive development to the field of art. We ran the same analysis for our second research question and found that 44.88\% of the survey participants agree that Generative AI art models are a positive development in the field of art. Notably, 27.02\% of artists selected 'neutral' on the Likert scale for this question. This analysis pertains to survey question number 14 (see Appendix A for survey questions). 

We conducted an ordinal logistic regression (OLR) analysis to investigate the association between Likert scale responses to RQ2 (dependent variable) and the secondary variables with a focus on the odds ratios obtained from the OLR model (as seen in Figure \ref{fig:RQ2fig}). Our analysis indicates a strong relationship (p-value  $<$ 0.05, all OLR detailed results can be seen in Appendix C) between agreeing that Generative AI art models are a positive development in the field of art and a subset of our secondary variables. Here, we see that 30-49 years olds, males, persons of color, participants who have used AI art models, and participants who said they would donate their artwork to train models were more likely to agree that generative AI models are a positive development for the field of art. Whereas makers (note that makers were a small n (n=5) with a fairly large range in the confidence interval bars represented in Figure \ref{fig:RQ2fig}), females, and those who identify as artists were less likely to agree that AI models are a positive development for the field of art.

\textbf{RQ3: What types of disclosure for the use of artwork to train an AI art model do artists consider fair?} 
\label{section:RQ3}

This research question investigates artists' perceptions of what types of disclosure are fair when using their artwork to train an AI art model. Our results show a consensus among a large majority of artists that transparency and disclosure regarding the specific images used in training an AI model are essential prerequisites for them to deem the model fair and acceptable. We see from our survey results that 80.17\% of the survey participants agree that creators of AI models should disclose the art and images used to train their models in detail. The analysis for RQ3 is related to survey question 13 (all survey questions can be seen in Appendix A), which we used to run an ordinal logistic regression (OLR) to explore the relationship between the Likert scale answers to RQ3 and our secondary variables. The results of the OLR, displayed in Figure \ref{fig:RQ3fig} as the odds ratio, show a strong relationship (p-value  $<$ 0.05, all odds ratios can be seen in Appendix C) between agreeing with the disclosure requirement and a subset of our secondary variables. 

Here, we see that participants with a general familiarity with AI art or were between the ages of 18-49 were more likely to agree that AI model creators should be required to disclose their training data in detail. Participants who expressed willingness to accept compensation in the form of a share of the derivatives or profits generated from AI artwork inspired by their work, along with a portion of the model creators' profits, also agreed that disclosure should be required. Participants who identified as illustrators (note that illustrators were a small n (n=9) with a fairly large range in the confidence interval bars represented in Figure \ref{fig:RQ3fig}) or were between the ages of 50-64 years old were less likely to agree that disclosure should be required.

\begin{figure*}
     \centering
         \includegraphics[width=0.85\columnwidth]{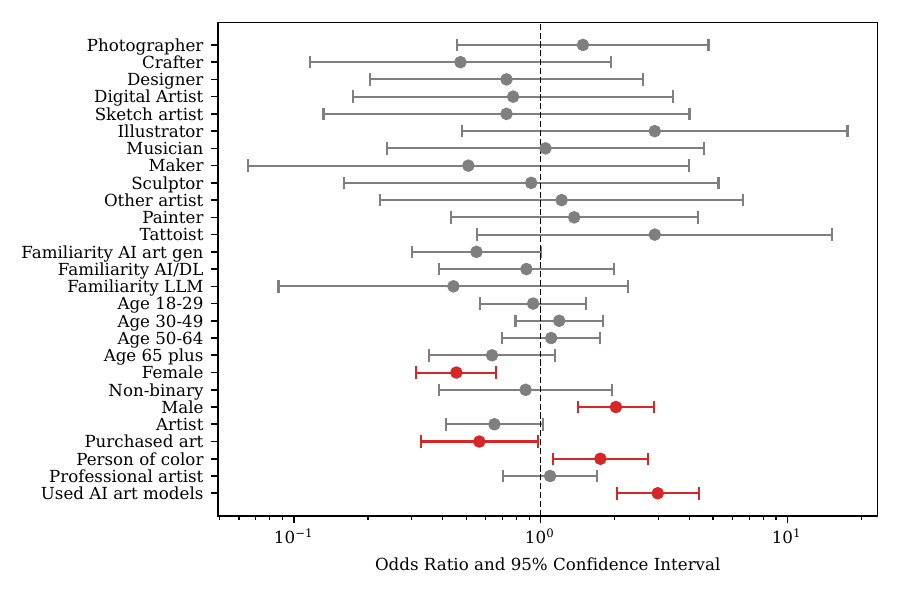}
         \includegraphics[width=0.85\columnwidth]{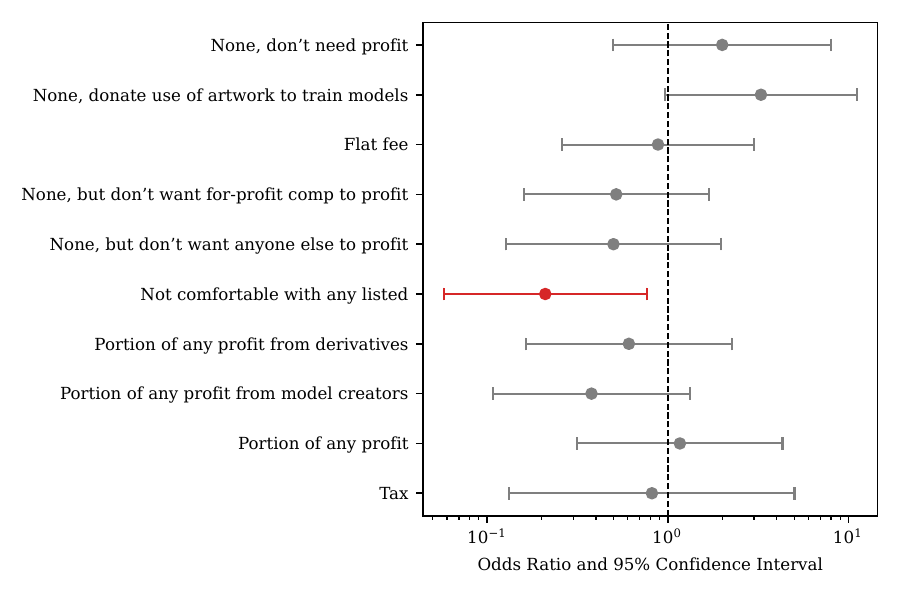}
        \caption{The ordinal logistic regression results show the association between answering `Agree' or `Strongly Agree' to RQ4A (If an AI art model was used by someone else to produce artwork recognizably in your style (e.g., in the style of Van Gogh) Should that work and its derivatives be considered the property of the person who used the AI model to generate the artwork?) and survey secondary variables. Results are represented by odds ratios on a log scale on the x-axis, with bars conveying a 95\% confidence interval and red indicating a p-value $<$ 0.05 and a confidence interval that does not cross the threshold of the dotted line in the figure.}
        \label{fig:RQ4_Afig}
\end{figure*}

\begin{figure}
     \centering
         \includegraphics[width=0.85\columnwidth]{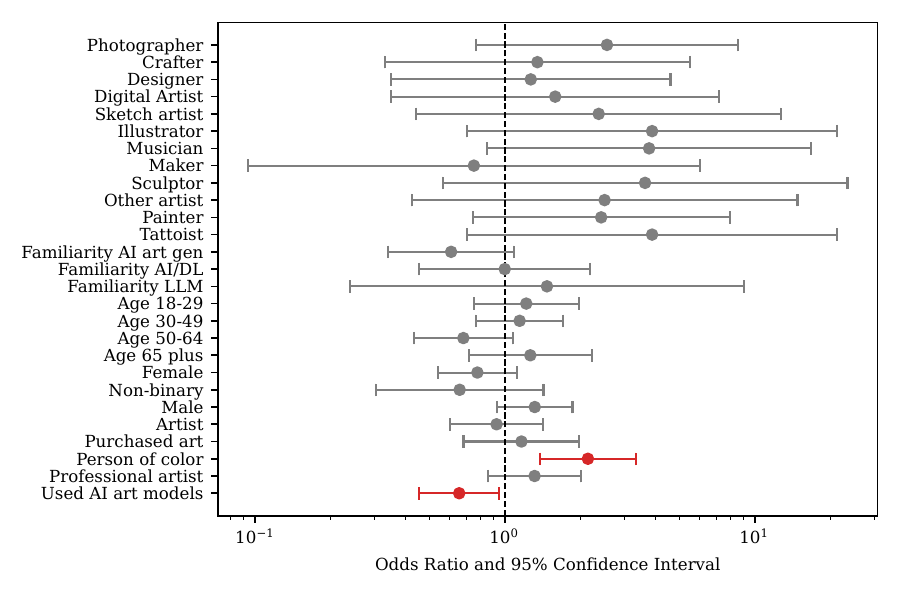}
         \includegraphics[width=0.85\columnwidth]{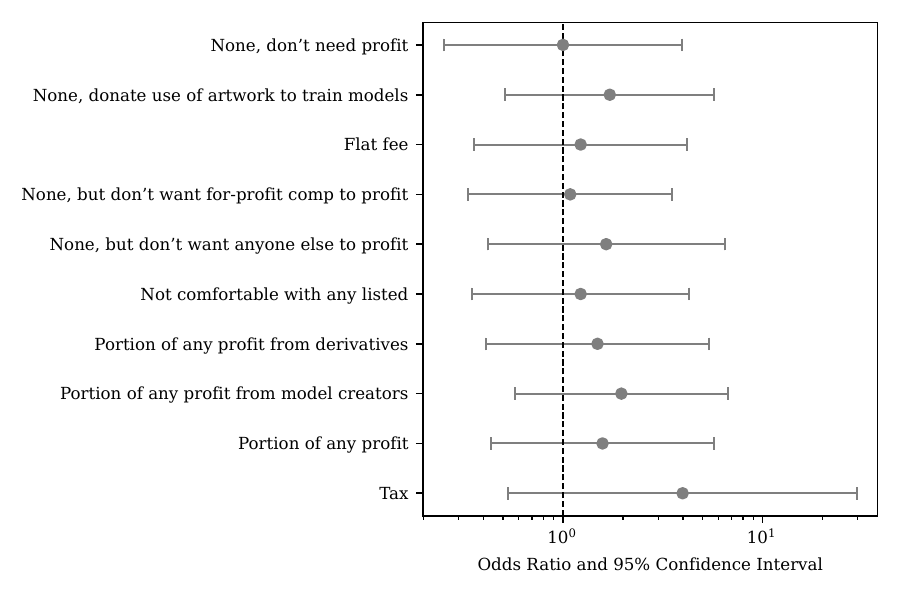}
        \caption{The ordinal logistic regression results show the association between answering `Agree' or `Strongly Agree' to RQ4B (If an AI art model was used by someone else to produce artwork recognizably in your style (e.g., in the style of Van Gogh and you are Van Gogh) Should that work and its derivatives be considered yours? and survey secondary variables. Results are represented by odds ratios on a log scale on the x-axis, with bars conveying a 95\% confidence interval and red indicating a p-value $<$ 0.05 and a confidence interval that does not cross the threshold of the dotted line in the figure.}
        \label{fig:RQ4_Bfig}
\end{figure}

\begin{figure}
     \centering
         \includegraphics[width=0.85\columnwidth]{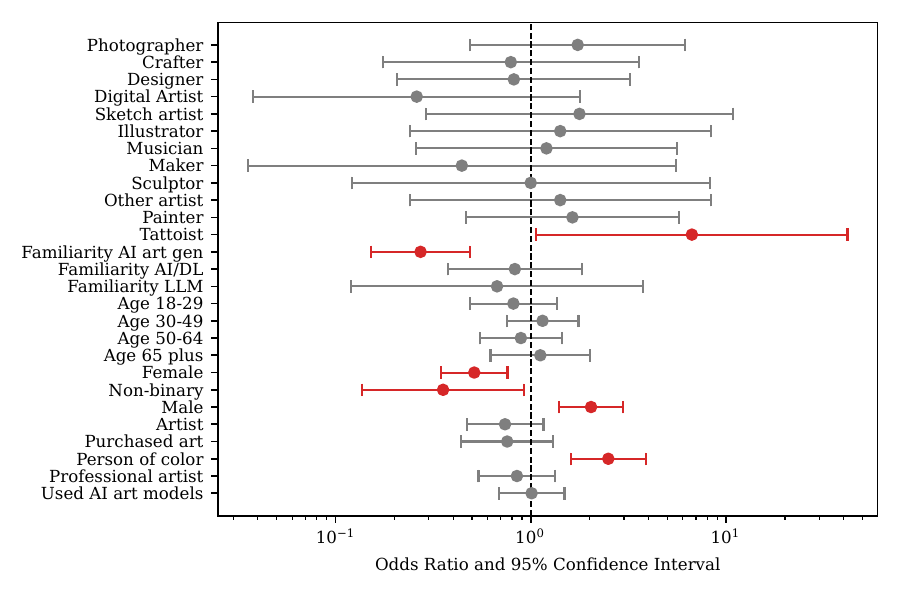}
         \includegraphics[width=0.85\columnwidth]{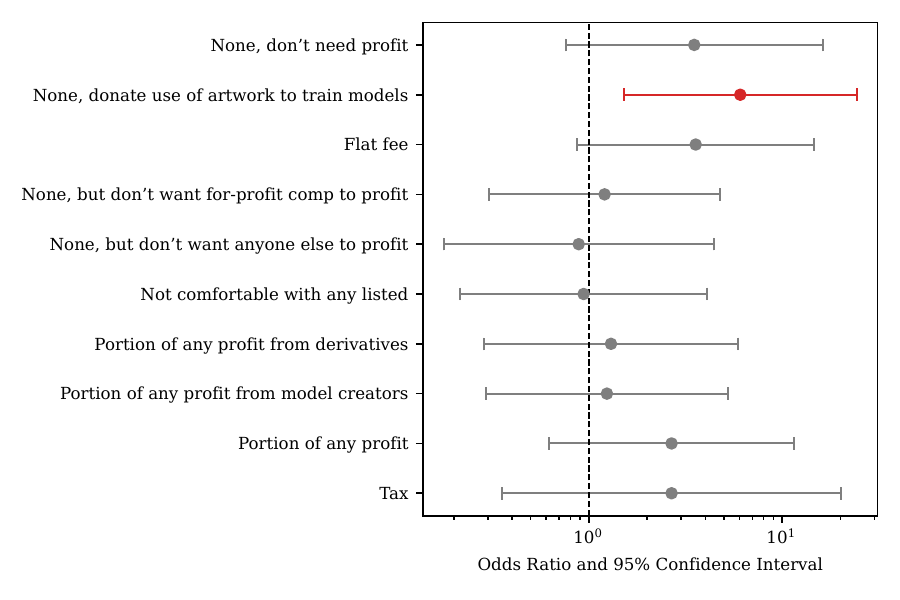}
        \caption{The ordinal logistic regression results show the association between answering `Agree' or `Strongly Agree' to RQ4C (If an AI art model was used by someone else to produce artwork recognizably in your style (e.g., in the style of Van Gogh) Should that work and its derivatives be considered the property of the AI model creators?) and survey secondary variables. Results are represented by odds ratios on a log scale on the x-axis, with bars conveying a 95\% confidence interval and red indicating a p-value $<$ 0.05 and a confidence interval that does not cross the threshold of the dotted line in the figure.}
        \label{fig:RQ4_Cfig}
\end{figure}

\textbf{RQ4: From an artist's perspective, who should own the derivative artwork created by AI art models?} 
\label{section:RQ4}

Finally, we look at the question that asks participants to think about a scenario where an AI art model was used by someone else to produce artwork recognizably in their (the survey participant's) style and think about who should own the work and its derivatives. The analysis for RQ4 is related to survey questions No. 8, 9, and 10 (all survey questions can be seen in Appendix A). 

\textbf{RQ4A: If an AI art model was used by someone else to produce artwork recognizably in your style (e.g., in the style of Van Gogh), Should that work and its derivatives be considered the property of the person who used the AI model to generate the artwork?} 

Here, we explore the scenario where the owner of the Generative AI artwork should be the person who used the AI model to generate the artwork. We see that 39.22\% of participants agreed that the work and its derivatives should be considered the property of the person who used the AI model to generate the artwork. We conducted an ordinal logistic regression (OLR) to investigate the relationship between our dependent variable (Likert scale responses to RQ4A) and a subset of our secondary variables, represented by the odds ratio depicted in Figure \ref{fig:RQ4_Afig}. 

We see evidence of a strong association (p-value  $<$ 0.05) between agreeing and our secondary variables. Namely, males, persons of color, and participants who have used Generative AI to make art were more likely to agree that the owner of the artwork in this scenario should be the person who used the AI model to generate the artwork. Whereas females, participants who had purchased a piece of artwork, and participants who said that they were not comfortable with any type of compensation for the use of their artwork to train a generative AI model were less likely to agree that the owner of the artwork in this scenario should be the person who used the AI model to generate the artwork.

\textbf{RQ4B: If an AI art model was used by someone else to produce artwork recognizably in your style (e.g., in the style of Van Gogh and you are Van Gogh), Should that work and its derivatives be considered yours?} 

Here, we explore the scenario where the owner of the Generative AI artwork should be the artist whose style is being recognizable and mimicked by the Generative AI art model. Our survey shows that 41.39\% of the participants agreed that the work and its derivatives should be considered their property (the artist whose style is being recognizable and mimicked by the Generative AI art model). 

To investigate the relationship between the Likert scale responses to RQ4B and the secondary variables, we conducted an ordinal logistic regression (OLR), presented in Figure \ref{fig:RQ4_Bfig} displaying the odds ratios derived from the analysis. Our results show that there is a strong relationship (p-value  $<$ 0.05, all detailed odds ratios can be seen in Appendix C) between agreeing that the artist whose style is recognizable in the output should be the owner of the art and a subset of our secondary variables. We see that participants who are persons of color are more likely to agree that the owner of the Generative AI artwork should be the artist whose style is recognizable and mimicked by the Generative AI art model. Whereas participants who have used Generative AI to make art were less likely to agree that the owner of the Generative AI artwork should be the artist whose style is recognizable and mimicked by the Generative AI art model.

\textbf{RQ4C: If an AI art model was used by someone else to produce artwork recognizably in your style (e.g., in the style of Van Gogh), Should that work and its derivatives be considered the property of the AI model creators?} 

Finally, we look at our final scenario where the owner of the Generative AI artwork should be the AI model creators. According to our survey results, only 26.80\% of the participants believed that the work and its derivatives should be considered the property of the creators of AI models. To examine the relationship between the answers to RQ4C on the Likert scale and other secondary variables, we conducted an ordinal logistic regression (OLR), as illustrated in Figure \ref{fig:RQ4_Cfig}. 

Our analysis suggests a strong relationship (p-value  $<$ 0.05, all odds ratios can be seen in Appendix C) between agreement on AI companies owning the resulting artwork and a subset of our secondary variables. Participants who considered themselves tattooists (note that tattooists were a small n (n=9) with a fairly large range in the confidence interval bars represented in Figure \ref{fig:RQ4_Cfig}), males, persons of color, or those who said they would donate their artwork to train generative AI models were more likely to agree that the owner of the Generative AI artwork should be the AI model creators. Whereas participants with a general familiarity with Generative AI Art, females, and non-binary participants were less likely to agree that the owner of the Generative AI artwork should be the AI model creators.

\section{Discussion}
\label{section:discussion}

The impact of Generative AI on artists is a complex global issue with consequences in many arenas, and it must be investigated through further multidisciplinary collaboration. We hope this work contributes to the existing literature on the ethics of AI-generated art and helps to provide a voice for artists on this topic.

In this study, we surveyed how artists feel about Generative Artificial Intelligence (AI) in art. Most participants believe that AI model creators should be required to disclose the art and images used to train them -- an aspect of Generative AI tools that is far from transparent. However, only a minority of participants see AI art models as a positive development in art, while the majority see them as a threat to art workers.

When asked who should own the work and its derivatives when an AI art model is used to produce artwork in a particular artist's style, the minority of participants believed that the AI model creators should own the work. In contrast, the majority of survey respondents were split between thinking it should belong to the person who used the AI model or the artist whose style is represented in the output.

Regarding compensation for using their artwork to train AI art models, over half of the participants did not require any compensation but were concerned about who was profiting from their artwork. One concern is that artists will be reluctant to share their work with each other and with the public going forward, and there may be resistance to perceived exploitation of that practice, especially in the service of profit. We think this interpretation is consistent with the result that some participants stated that they do not require profit but do not want for-profit companies to profit from their art. Others considered some compensation fair, while a minority felt taxing companies and individuals who profit from AI art outputs would be a fair option. Finally, some participants stated they would feel uncomfortable with the compensation options listed. Understanding how the field of Generative AI falls short in respecting the values of the people who make the cultural artifacts it relies upon -- and how that could be changed -- is an important ethical question that we hope to expand on in future work.

%-Future work
In our future work, we plan to increase the sample size and expand it to a more representative sample. This will increase the statistical power of our study and also allow us to look at how opinions change over time. Moreover, we would like to put the current flourishing of Generative AI within the context of art history and taxonomy. We believe that a longer timeline would help consolidate, unify, and clarify the relationship between technology and art. While it is essential to understand what artists think about Generative AI, we need to collaborate with experts from various fields, including art, history, philosophy, psychology, cognitive science, and others, to understand the implications of Generative AI on our understanding of art and its role in our lives. 

%All appendices and supplementary materials for this paper can be found on the arXiv version of this paper \cite{lovato2024foregrounding}. All data and all code are available from our GitHub repository, which can be found at the following URL: \texttt{https://github.com/juniperlovato/AASP}

\section{Acknowledgements} 

We are thankful to the anonymous reviewers for their insightful comments. Institutional Review Board Approval: The survey in this project is CHRBSS (Behavioral) STUDY00002379, approved by the University of Vermont I.R.B. on 3/14/2023. The authors would like to thank Maria Sckolnick, John Meluso, Kendall Fortney, Randall Harp, Laurent H\'ebert-Dufresne, Jean-Gabiel Young, and Winn Gillen, for their comments on early versions of this work. The Alfred P. Sloan Foundation award \#2024-22498, The National Science Foundation award \#2242829, and the MassMutual Center of Excellence in Complex Systems and Data Science support this work. Any opinions, findings, conclusions, or recommendations expressed in this material are those of the authors and do not necessarily reflect the views of the aforementioned financial supporters. 

\bibliography{aaai24}

\appendix

\section{Appendix A: Survey Questions}
\label{sec:survey_questions}

\begin{enumerate}
    \item Would you consider yourself to be an artist? (options: Yes; Somewhat; No; I'm not sure)
    \item What kind(s) of art do you practice? (select all that apply) (options: Photography; Paint; Drawing; Tattoo; Graphic Design; Illustration; Textile; Cinematography; Sculpture; Craft; Maker; Designer; Architecture; Jewelry; Writing; Music; Performance; Prefer not to respond; Not an artist but I work in an art-related field;	Not an artist, but I am an art enthusiast; Not an artist;  Other, please specify)
    \item Have you ever purchased a piece of art? (options: Yes; No; I'm not sure) 
    \item Do you consider yourself to be a professional artist? (options: Yes; No; Prefer not to respond) 
    \item Please read the following description of Artificial Intelligence (AI) art before continuing: AI art is any computer-generated artwork created through the use of artificial intelligence. There are several ways to generate AI art. Models used to generate AI art: Artificial intelligence algorithms use two or more images to train (training is a process used where an algorithm is fed data to learn from) a model to generate new AI art (e.g., Lensa, DeepDream). Large language models use artificial intelligence, which uses large amounts of text data to train a model to generate new AI language (e.g., GPT-3). Text-to-image models use textual prompts (e.g., “a yellow elephant”) from a user to generate AI art that depicts the textual prompt. The AI art is trained on large amounts of image and text data to train the model and generate new AI art (e.g., DALL-E, Imagen).  Example: Generated by DALL-E mini/Craiyon using the prompt “a cat in the style of Van Gogh.” According to its creators, “[t]he model was trained on unfiltered data from the Internet, limited to pictures with English descriptions.”
    \item Which method of AI art listed below had you heard of before starting this survey? (select all that apply) (options: AI art generally;	Artificial intelligence or deep learning algorithms, such as generative adversarial networks (GANs); Large Language Models; Text-to-image models: None of the above)
    \item Have you ever used AI to generate art? (options: Yes; No; I don't know) 
    \item If an AI art model was used by someone else to produce artwork recognizably in your style (e.g., in the style of Van Gogh), Should that work and its derivatives be considered the property of the person who used the AI model to generate the artwork? (options: Strongly agree; Agree; Neutral; Disagree; Strongly disagree)
    \item If an AI art model was used by someone else to produce artwork recognizably in your style (e.g., in the style of Van Gogh), Should that work and its derivatives be considered yours? (options: Strongly agree; Agree; Neutral; Disagree; Strongly disagree)
    \item If an AI art model was used by someone else to produce artwork recognizably in your style (e.g., in the style of Van Gogh), Should that work and its derivatives be considered the property of the AI model creators? (options: Strongly agree; Agree; Neutral; Disagree; Strongly disagree) 
    \item What type of compensation for the use of your artwork to train an AI art model would you consider fair? (select all that apply) (options: I would donate the use of my artwork to train AI art models (no compensation); I don't need to profit; I don’t need to profit, but I don't want for-profit-companies to profit from my art; I don’t need to profit, but I don’t want anyone else to profit from my art; A portion of any profit made; A portion of any profit made From the model creators; A portion of any profit from people who profit from derivative works made using the model; A flat fee; A flat fee from the model creators, for the use of my artwork as training data; flat fee for each derivative work sold (by anyone using the model); companies and individuals should be subject to a new tax by the government if they profit from AI art trained on other people’s work; I would not feel comfortable with any of the options listed above; Other, please specify) 
    \item You selected the response: Companies and individuals should be subject to a new tax in order to use AI art that was trained with your artwork. In your opinion, what should these taxes support? (option: open text field) 
    \item Do you think model creators should be required to disclose in detail what art and images they use to train their AI models? (options: Strongly agree; Agree; Neutral; Disagree; Strongly disagree)
    \item Do you agree with the following statement? "I see AI art models as a positive development in the field of art." (options: Strongly agree; Agree; Neutral; Disagree; Strongly disagree)
    \item Do you agree with the following statement? "I see AI models as a threat to art workers?" (options: Strongly agree; Agree; Neutral; Disagree; Strongly disagree)
    \item What is your age (options: 18-29; 30-49; 50-64, 65 or older) 
    \item Do you identify as a person of color? (options: Yes, No) 
    \item What is your gender identity? (options: Non-binary; Female; Male)
    \item In which country do you currently reside? (options: drop-down menu of all countries) 
\end{enumerate}

\section{Appendix B: Full Table of Descriptive Statistics}

A complete table of descriptive statistics for our participants in the survey can be found in Table \ref{tab:surveydescstats2}.

Descriptive statistics about the US artist workforce who are categorized as artists-related workers (US Census code 27-1010) can be seen in Table \ref{tab:artistsinusa}. Data comes from the 2020 US Census, US Census Bureau, and ACS PUMS 5-Year Estimate. Descriptive statistics about the U.S. artist workforce are categorized as artists-related workers (US Census code 27-1010), which can be seen in Table \ref{tab:artistsinusa}. Data comes from the 2020 US Census, US Census Bureau, and ACS PUMS 5-Year Estimate. The occupational categories include Art Directors, Craft Artists, Fine Artists, Painters, Sculptors, Illustrators, Special Effects Artists and Animators, Artists and Related Workers, All Other Designers, Commercial and Industrial Designers, Fashion Designers, Floral Designers, Graphic Designers, Interior Designers, Merchandise Displayers and Window Trimmers, Set and Exhibit Designers, Designers, All Other. This data demonstrates the number of US workers potentially implicated by Generative AI art. 

 \section{Appendix C: All Tables for Ordinal Logistic Regressions}

We ran an ordinal logistic regression (OLR) to explore the relationship between our dependent variable (the Likert scale answers to each primary research question). Our secondary variables, the results of these original logistic regressions, their odds-ratios, their p-values, and their confidence intervals (CI) can be seen in Tables \ref{tab:SM_T1}, \ref{tab:SM_T2}, \ref{tab:SM_T3}, \ref{tab:SM_T4}, \ref{tab:SM_T51}, \ref{tab:SM_T52}.  

\section{Appendix D: Contingency Tables}

All contingency tables for dependent variables with secondary variables can be seen in Tables \ref{tab:CT_1}, \ref{tab:CT_2},  \ref{tab:CT_3},  \ref{tab:CT_4},  \ref{tab:CT_5},  \ref{tab:CT_6},  \ref{tab:CT_7},  \ref{tab:CT_8},  \ref{tab:CT_9}, \ref{tab:CT_10}, \ref{tab:CT_11}, \ref{tab:CT_12}, and \ref{tab:CT_13}. 

\section{Appendix E: Tests for Independence}

\subsubsection{Kruskal-Wallis Test for Independence:} 

To test for independence between dependent variables that were ordinal with variables of interest that are categorical, we ran Kruskall-Wallis tests for independence. The results of these tests can be seen in Tables \ref{tab:KW_positive}, \ref{tab:KW_threat}, \ref{tab:KW_artist}, \ref{tab:KW_company}, \ref{tab:KW_disclosure}, \ref{tab:KW_user}. 

\textbf{Kruskal–Wallis Test}
The Kruskal-Wallis H-test is a non-parametric test that extends the Mann-Whitney U-test to compare differences between groups for an ordinal or continuous dependent variable. The explanatory groups may be ranked or unranked~\cite{statistics2015kruskal}. Its null hypothesis is stochastic homogeneity, which is equivalent to equality of the expected value of rank means, conditional on the homogeneity of rank variances~\cite{vargha1998kruskal}.

\subsubsection{Cochran-Armitage Test for Independence:} 

To test for independence between dependent variables that were ordinal with variables of interest that are binary, we ran Cochran-Armitage tests for independence. The results of these tests can be seen in Tables \ref{tab:CA_positive}, \ref{tab:CA_threat}, \ref{tab:CA_artist}, \ref{tab:CA_company}, \ref{tab:CA_disclosure}, and \ref{tab:CA_user}.   

\textbf{Cochran–Armitage Test}
The Cochran-Armitage test for trend uses weighted linear regression coefficients to test for trends in proportion between ordered levels of the response variable for a binary response variable. Its null hypothesis is a lack of trend, meaning that the proportions of each outcome are the same for each explanatory level~\cite{syntaxfreq}.
%\todo{add text that refs tables}

\begin{table}
\small
\centering
\tabcolsep=0.19cm
\begin{tabular}{@{}*{2}{c}@{}}
\textbf{Type} & \textbf{Number} \\ \hline
U.S. artist workforce & 144,000 \\
Avg. Age & 41 \\
Avg. Salary & \$57,958 \\
Avg Male Salary & 67,353 \\ 
Avg Female Salary & 45,618 \\ 
Male & 56.70\% \\ 
Female & 43.30\% \\
White (Non-Hispanic) & 73.20\% \\ 
Asian (Non-Hispanic) & 7.92\% \\
Black & 3.65\% \\
Two or more & 3.44\% \\ 
White (Hispanic) & 2.88\% \\  
American Indian & 0.01\% \\ 
Native Hawaiian & 0.13\% \\
Native other & 0.04\% \\
\hline
\end{tabular}
\caption{Descriptive statistics about the US artist workforce categorized as artists-related workers (US Census code 27-1010). Data comes from the 2020 US Census, US Census Bureau, and ACS PUMS 5-Year Estimate.}
\label{tab:artistsinusa}
\end{table}

\begin{table*} 
\small
\centering
\tabcolsep=0.19cm
\begin{tabular}{@{}*{2}{l}@{}}
\textbf{Type} & \textbf{Count \%} \\ \hline
Male & 273 (59.48\%)\\
Female  & 163 (35.51\%)\\
Non-binary & 23 (5.01\%)\\
White & 361 (78.65\%)\\
Person of Color & 98 (21.35\%)\\
Age 30-49 & 233 (50.76\%)\\
Age 18-29 & 104 (22.66\%)\\
Age 50-64 & 79 (17.21\%)\\
Age 65 or older & 43 (9.37\%)\\
Used AI Models - No  & 237 (49.17\%)\\
Used AI Models  - Yes & 231 (47.93\%)\\
Used AI Models - I don't know & 10 (2.07\%)\\
Used AI Models - None of the above & 4 (0.83\%)\\
Familiarity - AI art generally & 379 (78.63\%)\\
Familiarity -  AI/Deeplearning/GANS & 42 (8.71\%)\\
Familiarity -  Text-to-image & 12 (2.49\%)\\
Familiarity -  LLMS & 5 (1.04\%)\\
Familiarity - None & 44 (9.13\%)\\
Professional Artist - No & 256 (52.57\%)\\
Professional Artist - Yes & 214 (43.94\%)\\
Professional Artist - Prefer not to respond & 17 (3.49\%)\\
Artist - Yes & 310 (62.63\%)\\
Artist - Somewhat & 176 (35.56\%)\\
Artist - No & 7 (1.41\%)\\
Artist - I’m not sure & 2 (0.40\%)\\
Purchased Art - Yes & 422 (86.65\%)\\
Purchased Art - No & 59 (12.11\%)\\ 
Purchased Art - I’m not sure & 6 (1.23\%)\\
Art Practice - Photography & 116 (23.58\%)\\
Art Practice - Paint & 167 (33.94\%)\\
Art Practice - Writing & 52 (10.57\%)\\
Art Practice - Designer & 43 (8.74\%)\\
Art Practice - Craft & 20 (4.07\%)\\
Art Practice - Music & 18 (3.66\%)\\ 
Art Practice - Digital & 14(2.85\%) \\
Art Practice - Not artist & 13(2.64\%) \\
Art Practice - Other & 11(2.24\%) \\
Art Practice - Illustration & 9 (1.83\%)\\
Art Practice - Tattoo &	9 (1.83\%)\\
Art Practice - Drawing & 8 (1.63\%)\\
Art Practice - Sculpture & 7 (1.42\%)\\
Art Practice - Maker & 5 (1.02\%)\\ 
\hline
\end{tabular}
\caption{Descriptive statistics for our survey participants.}
\label{tab:surveydescstats2}
\end{table*}

\begin{table*} 
\small
\centering
\tabcolsep=0.19cm
\begin{tabular}{@{}*{6}{c}@{}}
\textbf{Variable} & \textbf{CI 2.5\%} & \textbf{97.5\%} & \textbf{Odds Ratio}  & \textbf{p-value} \\ \hline
Artist&	0.364481&	0.913316&	0.576963&	 $<$0.001\\
Purchased art&	0.331723&	1.004549&	0.577263&	 $<$0.001\\
Person of color&	1.23678&	3.085823&	1.953582&	 $<$0.001\\
Professional artist&	0.840948&	2.063429&	1.317284&	 $<$0.001\\
Used AI art models&	3.4805&	7.703421&	5.178007&	 $<$0.001\\
Male&	1.105324&	2.232347&	1.570817&	0.011789\\
Female&	0.417108&	0.868241&	0.601789&	 $<$0.001\\
Non-binary&	0.420314&	2.076117&	0.934141&	 $<$0.001\\
Age 50-64&	0.5506&	1.369427&	0.868336&	 $<$0.001\\
Age 65 plus&	0.346227&	1.0817&	0.611975&	 $<$0.001\\
Age 18-29&	0.56397&	1.493795&	0.917854&	0.730133\\
Age 30-49&	1.022423&	2.304733&	1.535061&	0.038743\\
Familiarity AI art gen&	0.594964&	1.937282&	1.073598&	0.813584\\
Familiarity AI/DL&	0.460722&	2.345589&	1.039551&	0.925566\\
Familiarity LLM	& 0.185243&	8.879363&	1.282512&	0.801008\\
Familiarity text to image&	0.256932&	3.219854&	0.90955&	0.883141\\
Photographer&	0.267789&	3.76309&	1.003849&	0.995454\\
Crafter&	0.059216&	1.147651&	0.260691&	0.075429\\
Designer&	0.203683&	3.386843&	0.830568&	0.795733\\
Digital Artist&	0.126946&	2.668446&	0.582021&	0.486019\\
Sketch artist&	0.222575&	9.567887&	1.459305&	0.693624\\
Illustrator&	0.100764&	3.509196&	0.594644&	0.566036\\
Musician&	0.11112&	2.489955&	0.526007&	0.417996\\
Maker&	0.005815&	0.967095&	0.074991&	0.047077\\
Sculptor&	0.016156&	1.091709&	0.132805&	0.060336\\
Other artist&	0.100764&	3.509204&	0.594644&	0.566036\\
Painter&	0.188397&	2.563246&	0.694917&	0.5847\\
Tattoist&	0.269396&	11.050416&	1.725382&	0.564828\\
Writer&	0.095046&	1.50597&	0.378334&	0.167879\\
None, don’t need profit&	0.779115&	12.922381&	3.173015&	0.107049\\
None, donate use of artwork to train models&	4.162129&	57.308125&	15.444217&	 $<$0.001\\
Flat fee&	0.445466&	5.012977&	1.49436&	0.515373\\
None, but don’t want for-profit comp to profit&	0.305542&	3.134003&	0.978555&	0.970882\\
None, but don’t want anyone else to profit&	0.387164&	5.704167	&1.486085&	0.563771\\
Not comfortable with any listed&	0.0935&	1.160233&	0.329365	& 0.083874\\
Portion of any profit from derivatives&	0.296462&	3.932296&	1.079711&	0.907418\\
Portion of any profit from model creators&	0.295023&	3.38909	&0.999929&	0.999909\\
Portion of any profit&	0.319872&	4.208076&	1.160191&	0.821179\\
Tax&	0.327069&	13.554205&	2.105507&	0.433234\\
\hline
\end{tabular}
\caption{Ordinal logistic regression (OLR). Dependent variable: agreeing or strongly agreeing that AI is a positive development for the field of art. Strong evidence follows a threshold of a p-value less than 0.05.}
\label{tab:SM_T1}
\end{table*}

\begin{table*} 
\small
\centering
\tabcolsep=0.19cm
\begin{tabular}{@{}*{6}{c}@{}}
\textbf{Variable} & \textbf{CI 2.5\%} & \textbf{CI 97.5\%} & \textbf{Odds Ratio}  & \textbf{p-value} \\ \hline
Artist&	0.673062&	1.734431&	1.080453&	 $<$0.001\\
Purchased art&	0.56515&	1.850283&	1.022589&	 $<$0.001\\
Person of color&	0.835831&	2.157276&	1.342802&	 $<$0.001\\
Professional artist&	0.511389&	1.27991&	0.809031&	 $<$0.001\\
Used AI art models&	0.381466&	0.843053&	0.567094&	 $<$0.001\\
Male&	0.68315&	1.452222&	0.996035&	 $<$0.001\\
Female&	0.678301&	1.484131&	1.003338&	 $<$0.001\\
Non-binary&	0.41613&	2.447633&	1.009224&	 $<$0.001\\
Age 50-64&	0.399295&	1.029979&	0.641299&	 $<$0.001\\
Age 65 plus&	0.385716&	1.286586&	0.704455&	 $<$0.001\\
Age 18-29&	1.337791&	4.055724&	2.329316&	 $<$0.001\\
Age 30-49&	0.829556&	1.938754&	1.26819&	 $<$0.001\\
Familiarity AI art gen&	0.55586&	2.015918&	1.058569&	0.862508\\
Familiarity AI/DL&	0.644924&	4.062339&	1.61861&	0.305025\\
Familiarity LLM&	0.160508&	6.218764&	0.999079&	0.999212\\
Familiarity text to image&	0.143725&	1.825538&	0.512226&	0.302196\\
Photographer&	0.624162&	8.7873&	2.341944&	0.207189\\
Crafter& 0.691218&	18.224995&	3.549288&	0.129129\\
Designer&	0.625025&	10.555753&	2.568582&	0.190796\\
Digital Artist&	0.349361&	7.951837&	1.666752&	0.521639\\
Sketch artist&	0.274778&	13.110306&	1.898007&	0.51577\\
Illustrator&	0.35758&	15.847783&	2.380514&	0.369873\\
Musician&	0.845339&	24.983415&	4.59559&	0.077486\\
Maker&	0.419635&	68.154064&	5.347879&	0.196624\\
Sculptor&	0.274778&	13.110325&	1.898006&	0.515771\\
Other artist&	0.603302&	34.971648&	4.593306&	0.141009\\
Painter&	0.716997&	9.797998&	2.650498&	0.14395\\
Tattoist&	0.603301&	34.971718&	4.593306&	0.141009\\
Writer&	0.58997&	9.564791&	2.375487&	0.223434\\
None, don’t need profit&	0.318227&	4.551266&	1.203467&	0.784937\\
None, donate use of artwork to train models &	0.366867 &	3.7479	& 1.172597 &	0.788264\\
Flat fee&	0.492379&	5.558458&	1.654347&	0.415569\\
None, but don’t want for-profit comp to profit&	0.672149&	6.774345&	2.133862&	0.198466\\
None, but don’t want anyone else to profit&	0.427385&	6.406989	&1.654767&	0.465872\\
Not comfortable with any listed&	0.880466&	11.318314&	3.156802	&0.077639\\
Portion of any profit from derivatives&	0.76601	& 11.308242	& 2.943165&	0.11599\\
Portion of any profit from model creators&	0.690259&	8.054443&	2.357891&	0.171143\\
Portion of any profit&	0.416284&	5.403195	&1.499755&	0.535396\\
Tax	&0.131044	&3.68014&	0.69445&	0.668239\\
\hline
\end{tabular}
\caption{Ordinal logistic regression (OLR). Dependent variable: agreeing or strongly agreeing that AI models are a threat to art workers. Strong evidence follows a threshold of a p-value less than 0.05.}
\label{tab:SM_T2}
\end{table*}

\begin{table*} 
\small
\centering
\tabcolsep=0.19cm
\begin{tabular}{@{}*{6}{c}@{}}
\textbf{Variable} & \textbf{CI 2.5\%} & \textbf{CI 97.5\%} & \textbf{Odds Ratio}  & \textbf{p-value} \\ \hline
Artist&	0.618355&	1.949844&	1.098041&	 $<$0.001\\
Purchased art&	0.851732&	3.164821&	1.641822&	 $<$0.001\\
Person of color&	0.809301&	2.740342&	1.489215&	 $<$0.001\\
Professional artist&	0.640443&	2.00734&	1.133837&	 $<$0.001\\
Used AI art models&	0.38645&	1.045068&	0.635505&	 $<$0.001\\
Male&	0.587018&	1.506886&	0.940515&	 $<$0.001\\
Female&	0.669859&	1.797992&	1.097452&	 $<$0.001\\
Non-binary&	0.30543&	2.406881&	0.857399&	 $<$0.001\\
Age 50-64&	0.236998&	0.712557&	0.410944&	 $<$0.001\\
Age 65 plus&	0.325902&	1.57366&	0.716142&	 $<$0.001\\
Age 18-29&	1.427261&	5.790101&	2.874715&	 $<$0.001\\
Age 30-49&	1.061875	&2.97578	&1.777612	&  $<$0.001\\
Familiarity AI art gen	&1.363085	&5.05368	&2.624614&	 $<$0.001\\
Familiarity AI/DL&	0.879414	&6.605245	&2.410134&	 $<$0.001\\
Familiarity text to image	&0.131881	&1.59315	&0.458373	&  $<$0.001\\
Photographer&	0.026665	&1.656513	&0.21017	&0.138655\\
Crafter&	0.020146&	1.971367	&0.199289&	0.167743\\
Designer	&0.021091	&1.532809	&0.179801	&0.116568\\
Digital Artist	&0.013077	&1.343422	&0.132543&	0.087241\\
Sketch artist	&0.018934	&6.359606	&0.347009	&0.475682\\
Illustrator	&0.006274	&0.725497	&0.067464	&0.026098\\
Musician	&0.022443	&2.549968	&0.239228	&0.23615\\
Other artist	&0.01451	&2.365917	&0.185285	&0.194532\\
Painter	&0.036576	&2.225627	&0.285316	&0.231452\\
Tattoist&	0.022214	&7.300981	&0.402718	&0.538408\\
Writer	&0.017016	&1.172106	&0.141224	&0.069849\\
None, don’t need profit	&0.105173	&1.407149	&0.384701	&0.14881\\
None, donate use of artwork to train models	&0.446605	&4.575144	&1.429434	&0.547225\\
Flat fee	&0.670956	&8.716645	&2.418365	&0.177032\\
None, but don’t want for-profit comp to profit	&0.639246	&6.378818	&2.019314&	0.231116\\
None, but don’t want anyone else to profit	&0.643891&	15.548169	&3.164068	&0.156186\\
Not comfortable with any listed	&0.559388	&7.355529&	2.028445	&0.281879\\
Portion of any profit from derivatives	&1.100951	&37.925791	&6.461768	&0.038783\\
Portion of any profit from model creators	&1.051359	&16.889636	&4.213914	&0.042288\\
Portion of any profit	&0.327584	&4.283127	&1.184518	&0.796243\\
\hline
\end{tabular}
\caption{Ordinal logistic regression (OLR). Dependent variable: agreeing or strongly agreeing that model creators should be required to disclose in detail what art \& images they use to train their AI models. Strong evidence follows a threshold of a p-value less than 0.05.}
\label{tab:SM_T3}
\end{table*}

\begin{table*} 
\small
\centering
\tabcolsep=0.19cm
\begin{tabular}{@{}*{6}{c}@{}}
\textbf{Variable} & \textbf{CI 2.5\%} & \textbf{CI 97.5\%} & \textbf{Odds Ratio}  & \textbf{p-value} \\ \hline
Artist&	0.468541&	1.160369&	0.737347&	0.187823\\
Purchased art&	0.439453&	1.302365&	0.756523&	0.314052\\
Person of color&	1.603519&	3.87071&	2.491336&	 $<$0.001\\
Professional artist&	0.539563&	1.334183&	0.848455&	0.476731\\
Used AI art models&	0.684392&	1.486553&	1.008655&	0.965265\\
Male&	1.394213&	2.965517&	2.033362&	 $<$0.001\\
Female&	0.34673&	0.759067&	0.513022&	 $<$0.001\\
Non-binary&	0.136972&	0.921049&	0.355187&	0.033244\\
Age 50-64&	0.547054&	1.443719&	0.888703&	0.633636\\
Age 65 plus&	0.621558&	2.010455&	1.117861&	0.709854\\
Age 18-29&	0.488074&	1.356759&	0.813756&	0.429419\\
Age 30-49&	0.750917&	1.753871&	1.147611&	0.524626\\
Familiarity AI art generally &	0.152611&	0.486134&	0.272377&	 $<$0.001\\
Familiarity AI/DL&	0.375901&	1.823414&	0.827903&	0.639206\\
Familiarity LLM&	0.120181&	3.748762&	0.671215&	0.649641\\
Familiarity text to image&	0.140257&	1.564695&	0.468465&	0.217808\\
Photographer&	0.487757&	6.18452&	1.73682&	0.394225\\
Crafter&	0.174597&	3.568943&	0.789385&	0.758671\\
Designer&	0.207425&	3.227255&	0.818176&	0.774409\\
Digital Artist&	0.037919&	1.787102&	0.260319&	0.170916\\
Sketch artist&	0.289616&	10.860348&	1.773507&	0.535461\\
Illustrator&	0.239963&	8.326074&	1.41349&	0.702103\\
Musician&	0.256817&	5.623486&	1.201752&	0.815437\\
Maker&	0.035642&	5.508235&	0.443085&	0.526711\\
Sculptor&	0.121091&	8.227352&	0.998126&	0.998609\\
Other artist&	0.239959&	8.326229&	1.41349&	0.702106\\
Painter&	0.4663&	5.715676&	1.632549&	0.443291\\
Tattoist&	1.065367&	41.810186&	6.674068&	0.042601\\
Writer&	0.088428&	1.500551&	0.364267&	0.16208\\
None, don’t need profit&	0.76008&	16.216441&	3.510811&	0.107708\\
None, donate use of artwork to train models&	1.512847&	24.393075	&6.074783&	0.01097\\
Flat fee&	0.871625&	14.630981&	3.571096&	0.076892\\
None, but don’t want for-profit comp to profit&	0.30342&	4.788926&	1.205429&	0.790657\\
None, but don’t want anyone else to profit&	0.177118&	4.423683&	0.885163&	0.881873\\
Not comfortable with any listed&	0.215123&	4.09741&	0.938854&	0.933113\\
Portion of any profit from derivatives&	0.285511&	5.930689&	1.30126&	0.733653\\
Portion of any profit from model creators&	0.292329&	5.254805	&1.239407&	0.770881\\
Portion of any profit&	0.621704&	11.541633&	2.678708&	0.186106\\
Tax&	0.355144&	20.211859&	2.679203&	0.339136\\
\hline
\end{tabular}
\caption{Ordinal logistic regression (OLR). Dependent variable: agreeing or strongly agreeing that if an AI art model was used by someone else to produce artwork recognizably in your style (e.g., in the style of Van Gogh) that work and its derivatives should be considered the property of the AI model creators. Strong evidence follows a threshold of a p-value less than 0.05.}
\label{tab:SM_T4}
\end{table*}

\begin{table*} 
\small
\centering
\tabcolsep=0.19cm
\begin{tabular}{@{}*{6}{c}@{}}
\textbf{Variable} & \textbf{CI 2.5\%} & \textbf{CI 97.5\%} & \textbf{Odds Ratio}  & \textbf{p-value} \\ \hline
Artist&	0.601364&	1.425504&	0.925876&	0.726499\\
Purchased art&	0.682946&	1.985811&	1.164561&	0.57583\\
Person of color&	1.382372&	3.336151&	2.147511&	 $<$0.001\\
Professional artist&	0.857238&	2.014212&	1.314024&	0.210158\\
Used AI art models&	0.453536&	0.949834&	0.656341&	0.025556\\
Male&	0.929958&	1.863216&	1.316326&	0.121057\\
Female&	0.540109&	1.114689&	0.775921&	 $<$0.001\\
Non-binary&	0.304406&	1.426784&	0.659031&	 $<$0.001\\
Age 50-64&	0.433393&	1.072792&	0.681865&	 $<$0.001\\
Age 65 plus&	0.715666&	2.230618&	1.263478&	 $<$0.001\\
Age 18-29&	0.749478&	1.976648&	1.21715&	0.427012\\
Age 30-49&	0.76602&	1.711638&	1.145054&	0.508996\\
Familiarity AI art gen&	0.341862&	1.085694&	0.609227&	0.092752\\
Familiarity AI/DL&	0.454393&	2.193558&	0.998367&	0.996753\\
Familiarity LLM&	0.239351&	9.063794&	1.472899&	0.676175\\
Familiarity text to image&	0.199754&	2.904406&	0.761687&	0.690169\\
Photographer&	0.767339&	8.542394&	2.560257&	0.126216\\
Crafter&	0.330235&	5.508344&	1.348721&	0.6769\\
Designer&	0.350222&	4.596068&	1.268718&	0.71705\\
Digital Artist&	0.350639&	7.202454&	1.58917&	0.547998\\
Sketch artist&	0.442624&	12.710739&	2.371937&	0.313265\\
Illustrator&	0.705023&	21.346164&	3.879373&	0.119181\\
Musician&	0.846183&	16.829236&	3.773675&	0.081682\\
Maker&	0.093845&	6.014805&	0.751304&	0.787605\\
Sculptor&	0.563212&	23.470174&	3.635751&	0.174907\\
Other artist&	0.423624&	14.806303&	2.504456&	0.311247\\
Painter&	0.742785&	7.928239&	2.426721&	0.142187\\
Tattoist&	0.705023&	21.346176&	3.879373&	0.119182\\
Writer&	0.37674&	4.75844&	1.338916&	0.65191\\
None, don’t need profit&	0.25232&	3.961372&	0.999766&	0.999734\\
None, donate use of artwork to train models&	0.513752&	5.733981&	1.716346&	0.380072\\
Flat fee&	0.358686&	4.194022&	1.226515&	0.744814\\
None, but don’t want for-profit comp to profit&	0.334713&	3.531268	&1.08718&	0.889399\\
None, but don’t want anyone else to profit&	0.418543&	6.478014&	1.646611&	0.475448\\
Not comfortable with any listed&	0.349148&	4.307525&	1.226361&	0.750228\\
Portion of any profit from derivatives&	0.409981&	5.41326&	1.489744&	0.544844\\
Portion of any profit from model creators&	0.573956&	6.71113	&1.962624&	0.282421\\
Portion of any profit&	0.435868&	5.718306&	1.578742&	0.486822\\
Tax	&0.532577&	29.762374&	3.981302&	0.178264\\
\hline
\end{tabular}
\caption{Ordinal logistic regression (OLR). Dependent variable: agreeing or strongly agreeing that if an AI art model was used by someone else to produce artwork recognizably in your style (e.g., in the style of Van Gogh and you are Van Gogh) that work and its derivatives should be considered the property of the artist. Strong evidence follows a threshold of a p-value less than 0.05.}
\label{tab:SM_T51}
\end{table*}

\begin{table*} 
\small
\centering
\tabcolsep=0.19cm
\begin{tabular}{@{}*{6}{c}@{}}
\textbf{Variable} & \textbf{CI 2.5\%} & \textbf{CI 97.5\%} & \textbf{Odds Ratio}  & \textbf{p-value} \\ \hline
Artist&	0.414702&	1.019737&	0.650297&	 $<$0.001\\
Purchased art&	0.327522&	0.974205&	0.564866&	 $<$0.001\\
Person of color&	1.120118&	2.725496&	1.747248&	 $<$0.001\\
Professional artist&	0.702605&	1.699884&	1.092862&	 $<$0.001\\
Used AI art models&	2.034051&	4.377637&	2.984013&	 $<$0.001\\
Male&	1.41509&	2.883435&	2.01998&	 $<$0.001\\
Female&	0.314178&	0.66174&	0.455965&	 $<$0.001\\
Non-binary&	0.389118&	1.943853&	0.869706&	 $<$0.001\\
Age 50-64&	0.698376&	1.745508&	1.104093&	0.671752\\
Age 65 plus&	0.352683&	1.146873&	0.63599&	0.132472\\
Age 18-29&	0.570787&	1.525004&	0.93298&	0.782005\\
Age 30-49&	0.791308&	1.788877&	1.18977&	0.40368\\
Familiarity AI art generally&	0.301199&	1.004236&	0.549977&	0.05163\\
Familiarity AI/DL&	0.388081&	1.979027&	0.876369&	0.750841\\
Familiarity LLM&	0.086738&	2.268525&	0.443584&	0.328954\\
Familiarity text to image&	0.221942&	2.714057&	0.776121&	0.691518\\
Photographer&	0.460161&	4.790509&	1.484723&	0.50843\\
Crafter&	0.116186&	1.930897&	0.473649&	0.297291\\
Designer&	0.204141&	2.595388&	0.727891&	0.624392\\
Digital Artist&	0.174549&	3.434168&	0.77423&	0.736364\\
Sketch artist&	0.132029&	4.012481&	0.727848&	0.715319\\
Illustrator&	0.480708&	17.524118&	2.902409&	0.245438\\
Musician&	0.237822&	4.606483&	1.046672&	0.95189\\
Maker&	0.065305&	3.981952&	0.509941&	0.520716\\
Sculptor&	0.15958&	5.25934&	0.916126&	0.921737\\
Other artist&	0.224562&	6.596202&	1.21707&	0.819786\\
Painter&	0.432178&	4.332419&	1.368348&	0.593818\\
Tattoist&	0.554963&	15.175121&	2.902004&	0.206845\\
Writer&	0.230251&	2.832793&	0.807621&	0.738605\\
None, don’t need profit&	0.498665&	8.015598&	1.999275&	0.328153\\
None, donate use of artwork to train models&	0.965801&	11.105517&	3.275014&	0.056896\\
Flat fee&	0.259222&	3.002946&	0.882286&	0.841163\\
None, but don’t want for-profit comp to profit&	0.159296&	1.686612&	0.518334&	0.274998\\
None, but don’t want anyone else to profit&	0.126867&	1.973433	&0.500362&	0.322656\\
Not comfortable with any listed&	0.05761&	0.765249&	0.209967	&0.018008\\
Portion of any profit from derivatives&	0.164151&	2.258827&	0.608925&	0.458287\\
Portion of any profit from model creators&	0.108502&	1.319073	&0.378315&	0.127163\\
Portion of any profit&	0.315746&	4.308758&	1.166393&	0.817425\\
Tax&	0.132833&	5.021371&	0.816703&	0.827032\\
\hline
\end{tabular}
\caption{Ordinal logistic regression (OLR). Dependent variable: agreeing or strongly agreeing that if an AI art model was used by someone else to produce artwork recognizably in your style (e.g., in the style of Van Gogh) that work and its derivatives should be considered the property of the person who used the AI model to generate the artwork. Strong evidence follows a threshold of a p-value less than 0.05.}
\label{tab:SM_T52}
\end{table*}

\begin{table*}
\small
\centering
\tabcolsep=0.19cm
\begin{tabular}{@{}*{3}{c}@{}}
\textbf{Secondary Variable (Binary)} & \textbf{Dependent Variable (Likert)}& \textbf{Contingency Table [SD D N A SA]} \\ \hline
Not Person of color & AI Positive for Art & [[50 53 90 76 66] \\
Person of color & AI Positive for Art  & [ 6 11 22 24 29]]\\
Not Person of color & AI Threat to Art	 & [[ 22  53  54 113  93]\\
Person of color & AI Threat to Art  & [  6  10  14  32  30]]\\
Not Person of color & Required Disclosure  & [[  9  21  38  97 170]\\
Person of color & Required Disclosure  & [  3   2  10  37  40]]\\
Not Person of color & Owner AI Company  & [[124  79  59  41  32]\\
Person of color & Owner AI Company &[ 18  19  14  23  18]]\\
Not Person of color & Owner AI User &[[75 67 73 61 59]\\
Person of color & Owner AI User  & [ 8 23 13 27 21]]\\
Not Person of color & Owner Artist &[[36 94 81 68 56]\\
Person of color & Owner Artist  &[ 5 17 18 24 28]]\\
\hline
\end{tabular}
\caption{Contingency table for the secondary variable of interest (person of color) and dependent variables on a 5-point Likert scale strongly disagree [(SD, leftmost number in bracket), disagree (D), neutral (N), agree (A), and strongly agree (SA, rightmost number in bracket)]. Outer brackets indicate the start and end of each contingency table, and the inner brackets indicate the start and end of a row in the table.}
\label{tab:CT_1}
\end{table*}

\begin{table*}
\small
\centering
\tabcolsep=0.19cm
\begin{tabular}{@{}*{3}{c}@{}}
\textbf{Secondary Variable (Binary)} & \textbf{Dependent Variable (Likert)} & \textbf{Contingency Table [SD D N A SA]} \\ \hline
Not Professional Artist & AI Positive for Art & [[33 34 63 60 39]\\
Professional Artist & AI Positive for Art & [23 30 49 40 56]]\\
Not Professional Artist & AI Threat to Art & [[12 31 35 89 62]\\
Professional Artist & AI Threat to Art  &[16 32 33 56 61]]\\
Not Professional Artist & Required Disclosure & [[  5  11  31  89  93]\\
Professional Artist & Required Disclosure & [7  12  17  45 117]]\\
Not Professional Artist & Owner AI Company &[[59 56 47 45 22]\\
Professional Artist &Owner AI Company & [83 42 26 19 28]]\\
Not Professional Artist	& Owner AI User	& [[37 52 53 52 35]\\
Professional Artist	& Owner AI User	& [46 38 33 36 45]]\\
Not Professional Artist	& Owner Artist &  [[15 67 59 50 38]\\
Professional Artist	& Owner Artist & [26 44 40 42 46]]\\
\hline
\end{tabular}
\caption{Contingency table for the secondary variable of interest (professional artist) and dependent variables on a 5-point Likert scale strongly disagree [(SD, leftmost number in bracket), disagree (D), neutral (N), agree (A), and strongly agree (SA, rightmost number in bracket)]. Outer brackets indicate the start and end of each contingency table, and the inner brackets indicate the start and end of a row in the table.}
\label{tab:CT_2}
\end{table*}

\begin{table*}
\small
\centering
\tabcolsep=0.19cm
\begin{tabular}{@{}*{3}{c}@{}}
\textbf{Secondary Variable (Binary)} & \textbf{Dependent Variable (Likert)} &  \textbf{Contingency Table [SD D N A SA]} \\ \hline
Not Purchased Art &	AI Positive for Art	& [[ 7  3 17 12 10]\\
Purchased Art &	AI Positive for Art	& [49 61 95 88 85]]\\
Not Purchased Art &	AI Threat to Art & [[  1   6   9  21  12]\\
Purchased Art &	AI Threat to Art & [27  57  59 124 111]]\\
Not Purchased Art &	Required Disclosure	& [[  0   1  13  19  16]\\
Purchased Art &	Required Disclosure	& [12  22  35 115 194]]\\
Not Purchased Art & Owner AI Company & [[ 10  12  14   7   6]\\
Purchased Art & Owner AI Company & [132  86  59  57  44]]\\
Not Purchased Art &	Owner AI User & [[ 4 11 12 13  9]\\
Purchased Art &	Owner AI User & [79 79 74 75 71]]\\
Not Purchased Art &	Owner Artist & [[ 4 13 13 12  7]\\
Purchased Art &	Owner Artist & [37 98 86 80 77]]\\
\hline
\end{tabular}
\caption{Contingency table for the secondary variable of interest (purchased art) and dependent variables on a 5-point Likert scale strongly disagree [(SD, leftmost number in bracket), disagree (D), neutral (N), agree (A), and strongly agree (SA, rightmost number in bracket)]. Outer brackets indicate the start and end of each contingency table, and the inner brackets indicate the start and end of a row in the table.}
\label{tab:CT_3}
\end{table*}

\begin{table*} 
\small
\centering
\tabcolsep=0.19cm
\begin{tabular}{@{}*{3}{c}@{}}
\textbf{Secondary Variable (Binary)} & \textbf{Dependent Variable (Likert)} & \textbf{Contingency Table [SD D N A SA]} \\ \hline
Not Used AI art	& AI Positive for Art & [[44 45 65 38 19]\\
Used AI art	& AI Positive for Art & [12 19 47 62 76]]\\
Not Used AI art	AI & Threat to Art & [[ 6 24 35 76 70]\\
Used AI art	AI & Threat to Art & [22 39 33 69 53]]\\
Not Used AI art	& Required Disclosure & [[  4   8  24  64 111]\\
Used AI art	& Required Disclosure & [  8  15  24  70  99]]\\
Not Used AI art	& Owner AI Company & [[58 54 47 36 16]\\
Used AI art	& Owner AI Company & [84 44 26 28 34]]\\
Not Used AI art	& Owner AI User	& [[49 58 45 29 30]\\
Used AI art	& Owner AI User	& [34 32 41 59 50]]\\
Not Used AI art &	Owner Artist &	[[17 46 56 48 44]\\
Used AI art &	Owner Artist & [24 65 43 44 40]]\\
\hline
\end{tabular}
\caption{Contingency table for the secondary variable of interest (used AI art) and dependent variables on a 5-point Likert scale strongly disagree [(SD, leftmost number in bracket), disagree (D), neutral (N), agree (A), and strongly agree (SA, rightmost number in bracket)]. Outer brackets indicate the start and end of each contingency table, and the inner brackets indicate the start and end of a row in the table.}
\label{tab:CT_4}
\end{table*}

\begin{table*} 
\small
\centering
\tabcolsep=0.19cm
\begin{tabular}{@{}*{3}{c}@{}}
\textbf{Secondary Variable (binary)} & \textbf{Dependent Variable (Likert)} & \textbf{Contingency Table [SD D N A SA]} \\ \hline
Not Artist & Owner AI User & [[  1   0   4   1   0]\\
Artist & Owner AI User & [88 102  94  95  84]]\\
Not Artist &	Owner Artist & [[ 0   5   0   0   1]\\
Artist &	Owner Artist & [43 116 115 101  87]]\\
Not Artist	& Owner AI Company	& [[  3   0   2   1   0]\\
Artist	& Owner AI Company	&  [145 113  84  68  53]]\\
Not Artist & Required Disclosure & [[  0   0   0   3   2]\\
Artist & Required Disclosure & [13  24  55 143 218]]\\
Not Artist	& AI Positive for Art & [[  1   0   1   3   0]\\
Artist	& AI Positive for Art & [59  69 123 101 100]]\\
Not Artist	& AI Threat to Art	& [[  0   1   2   1   1]\\
Artist	& AI Threat to Art	& [28  64  78 154 128]]\\
\hline
\end{tabular}
\caption{Contingency table for the secondary variable of interest (Artist) and dependent variables on a 5-point Likert scale strongly disagree [(SD, leftmost number in bracket), disagree (D), neutral (N), agree (A), and strongly agree (SA, rightmost number in bracket)]. Outer brackets indicate the start and end of each contingency table, and the inner brackets indicate the start and end of a row in the table.}
\label{tab:CT_9}
\end{table*}

\begin{table*} 
\small
\centering
\tabcolsep=0.19cm
\begin{tabular}{@{}*{3}{c}@{}}
\textbf{Secondary Variable (Categorical)} & \textbf{Dependent Variable (Likert)} & \textbf{Contingency Table [SD D N A SA]} \\ \hline
Age	18-29 & Owner AI User & [[42 46 46 57 40]\\
Age 30-49 & Owner AI User &[24 24 17 22 17]\\
Age	50-64 & Owner AI User	& [16 12 19 11 21]\\
Age	65 plus & Owner AI User	& [ 5 16 12  4  6]]\\
Age	18-29 & Owner Artist & [[21 62 49 61 38]\\
Age	30-49 & Owner Artist & [10 27 20 18 29]\\
Age	50-64 & Owner Artist & [10 21 24 10 14]\\
Age	65 plus & Owner Artist & [ 2  6 17 11  6]]\\
Age	18-29 & Owner AI Company & [[77 47 36 38 33]\\
Age	30-49 & Owner AI Company &  [33 30 19 12 10]\\
Age50-64	& Owner AI Company &  [29 16 14 12  8]\\
Age	65 plus & Owner AI Company &  [ 6 15 13  7  2]]\\
Age	18-29 & Required Disclosure & [[  6  13  25  75 112]\\
Age	30-49 & Required Disclosure & [  1   2  10  38  53]\\
Age	50-64 & Required Disclosure & [  6   5  14  19  35]\\
Age	65 plus & Required Disclosure & [  0   4   5  14  20]]\\
Age	18-29 & AI Positive for Art & [[22 33 60 59 56]\\
Age	30-49 & AI Positive for Art & [20 19 22 20 23]\\
Age	50-64& AI Positive for Art & [12  9 25 15 18]\\
Age	65 plus & AI Positive for Art & [ 6  8 17  9  3]]\\
Age 18-29 &	AI Threat to Art &	[[16 37 35 78 65]\\
Age 30-49 &	AI Threat to Art & [ 7  9 10 44 34]\\
Age 50-64&	AI Threat to Art & [ 4 13 21 22 18]\\
Age 65 plus&	AI Threat to Art & [ 1  6 13 11 12]]\\
\hline
\end{tabular}
\caption{Contingency table for the secondary variable of interest (Age) and dependent variables on a 5-point Likert scale strongly disagree [(SD, leftmost number in bracket), disagree (D), neutral (N), agree (A), and strongly agree (SA, rightmost number in bracket)]. Outer brackets indicate the start and end of each contingency table, and the inner brackets indicate the start and end of a row in the table.}
\label{tab:CT_5}
\end{table*}

\begin{table*} 
\small
\centering
\tabcolsep=0.19cm
\begin{tabular}{@{}*{3}{c}@{}}
\textbf{Secondary Variable (Categorical)} & \textbf{Dependent Variable (Likert)} & \textbf{Contingency Table [SD D N A SA]} \\ \hline
AI Model Familiarity: general &	Owner AI User & [[82 81 67 73 67]\\
AI Model Familiarity: deep learning & Owner AI User &  [ 2  8 14  9  8]\\
AI Model Familiarity: large language models &	Owner AI User & [ 5  8 11 11  7]\\
AI Model Familiarity: text-to-image & Owner AI User & [ 0  3  4  3  1]\\
AI Model Familiarity: none &	Owner AI User & [ 0  2  2  0  1]]\\

AI Model Familiarity: general &	Owner Artist & [[ 40 106  76  80  68]\\ 
AI Model Familiarity: deep learning & Owner Artist & [  1   3  20   8   8]\\
AI Model Familiarity: large language models & Owner Artist &  [  2   7  16   8   9]\\
AI Model Familiarity: text-to-image & Owner Artist &  [  0   4   2   2   3]\\
AI Model Familiarity: none & Owner Artist &  [  0   1   1   3   0]]\\

AI Model Familiarity: general & Owner AI Company & [[137  97  49  46  41] \\ 
AI Model Familiarity: deep learning  & Owner AI Company & [  3   3  20  11   4]\\
AI Model Familiarity: large language models & Owner AI Company & [  7   8  11  10   6]\\
AI Model Familiarity: text-to-image & Owner AI Company & [  1   3   5   1   1]\\
AI Model Familiarity: none  & Owner AI Company & [  0   2   1   1   1]]\\

AI Model Familiarity: general & Required Disclosure	& [[ 10  18  34 109 193]\\
AI Model Familiarity: deep learning  & Required Disclosure & [  1   3  12  14  10]\\
AI Model Familiarity: large language models  & Required Disclosure &  [  1   2   5  18  13]\\
AI Model Familiarity: text-to-image & Required Disclosure &  [  1   1   4   2   2]\\
AI Model Familiarity: none & Required Disclosure &  [  0   0   0   3   2]]\\

AI Model Familiarity: general & AI Positive for Art	& [[52 51 94 82 84]\\
AI Model Familiarity: deep learning  & AI Positive for Art &  [ 4  5 17  9  5]\\
AI Model Familiarity: large language models  & AI Positive for Art &  [ 4  8 10 10  7]\\
AI Model Familiarity: text-to-image & AI Positive for Art &  [ 0  3  3  1  3]\\
AI Model Familiarity: none  & AI Positive for Art &  [ 0  2  0  2  1]]\\

AI Model Familiarity: general & AI Threat to Art &[[ 26  51  59 117 110]\\
AI Model Familiarity: deep learning  & AI Threat to Art & [  0   6  11  17   6]\\
AI Model Familiarity: large language models  & AI Threat to Art & [  1   5   6  16  11]\\
AI Model Familiarity: text-to-image & AI Threat to Art &  [  1   2   3   3   1]\\
AI Model Familiarity: none & AI Threat to Art &  [  0   1   1   2   1]]\\
\hline
\end{tabular}
\caption{Contingency table for the secondary variable of interest (AI model familiarity) and dependent variables on a 5-point Likert scale strongly disagree [(SD, leftmost number in bracket), disagree (D), neutral (N), agree (A), and strongly agree (SA, rightmost number in bracket)]. Outer brackets indicate the start and end of each contingency table, and the inner brackets indicate the start and end of a row in the table.}
\label{tab:CT_6}
\end{table*}

\begin{table*} 
\small
\centering
\tabcolsep=0.19cm
\begin{tabular}{@{}*{3}{c}@{}}
\textbf{Secondary Variable (Categorical)} & \textbf{Dependent Variable (Likert)} & \textbf{Contingency Table [SD D N A SA]} \\ \hline
Art Practice: Paint &	Owner AI User & [[28 30 36 41 28]\\
Art Practice: Photography &	Owner AI User &  [11 29 22 27 23]\\
Art Practice: Writing &	Owner AI User &  [14 12  6  6 11]\\
Art Practice: Designer &	Owner AI User &  [11 12  4  7  7]\\
Art Practice: Craft &	Owner AI User &  [ 4  6  7  1  0]\\
Art Practice: Music &	Owner AI User &  [ 5  4  1  3  4]\\
Art Practice: Digital &	Owner AI User &  [ 4  3  3  2  2]\\
Art Practice: Not Artist &	Owner AI User &  [ 2  2  5  1  1]\\
Art Practice: Other &	Owner AI User &  [ 4  0  2  4  0]\\
Art Practice: Tattoo &	Owner AI User &  [ 0  0  4  2  2]\\
Art Practice: Illustration &	Owner AI User &  [ 3  0  0  2  4]\\
Art Practice: Drawing &	Owner AI User &  [ 0  3  3  0  1]\\
Art Practice: Sculpture &	Owner AI User &  [ 1  1  3  0  1]\\
Art Practice Maker & Owner AI User &  [ 2  0  2  0  0]]\\
 
Art Practice: Paint &	Owner Artist &	[[13 38 41 37 34]\\
Art Practice: Photography &	 Owner Artist &[ 6 30 22 30 23]\\
Art Practice: Writing & Owner Artist &[ 7 15 12  8  7]\\
Art Practice: Designer & Owner Artist &[ 9 12  8  7  5]\\
Art Practice: Craft & Owner Artist &[ 1  6  7  2  2]\\
Art Practice: Music & Owner Artist &[ 1  3  4  7  2]\\
Art Practice: Digital & Owner Artist &[ 1  5  3  4  1]\\
Art Practice: Not Artist & Owner Artist &[ 0  6  4  0  1]\\
Art Practice: Other & Owner Artist &[ 2  2  1  3  2]\\
Art Practice: Tattoo & Owner Artist &[ 0  1  3  0  4]\\
Art Practice: Illustration & Owner Artist &[ 1  1  2  1  4]\\
Art Practice: Drawing & Owner Artist &[ 0  1  4  1  1]\\
Art Practice: Sculpture & Owner Artist &[ 0  1  2  1  2]\\
Art Practice Maker & Owner Artist &[ 2  0  2  0  0]]\\
 
Art Practice: Paint &	Owner AI Company & [[42 39 29 28 25]\\
Art Practice: Photography & Owner AI Company &[30 25 20 22 15]\\
Art Practice: Writing & Owner AI Company &[25 14  7  2  1]\\
Art Practice: Designer & Owner AI Company & [17 10  6  5  3]\\
Art Practice: Craft & Owner AI Company &[ 6  5  6  1  0]\\
Art Practice: Music & Owner AI Company &[ 3  7  3  2  2]\\
Art Practice: Digital & Owner AI Company &[ 6  6  1  1  0]\\
Art Practice: Not Artist & Owner AI Company & [ 4  2  4  1  0]\\
Art Practice: Other & Owner AI Company &[ 4  0  4  2  0]\\
Art Practice: Tattoo & Owner AI Company &[ 1  0  2  1  4]\\
Art Practice: Illustration & Owner AI Company &[ 3  2  1  1  2]\\
Art Practice: Drawing & Owner AI Company &[ 1  2  2  2  0]\\
Art Practice: Sculpture & Owner AI Company &[ 4  0  0  1  1]\\
Art Practice Maker & Owner AI Company &[ 2  1  1  0  0]]\\
\hline
\end{tabular}
\caption{Contingency table for the secondary variable of interest (Art Practice) and dependent variables on a 5-point Likert scale strongly disagree [(SD, leftmost number in bracket), disagree (D), neutral (N), agree (A), and strongly agree (SA, rightmost number in bracket)]. Outer brackets indicate the start and end of each contingency table, and the inner brackets indicate the start and end of a row in the table.}
\label{tab:CT_7}
\end{table*}

\begin{table*} 
\small
\centering
\tabcolsep=0.19cm
\begin{tabular}{@{}*{3}{c}@{}}
\textbf{Secondary Variable (Categorical)} & \textbf{Dependent Variable (Likert)} & \textbf{Contingency Table [SD D N A SA]} \\ \hline
Art Practice: Paint &	Required Disclosure	& [[ 2  5 20 64 71]\\
Art Practice: Photography & Required Disclosure	& [ 2  7 13 28 59]\\
Art Practice: Writing & Required Disclosure	& [ 5  3  4 14 20]\\
Art Practice: Designer & Required Disclosure & [ 3  3  3  9 22]\\
Art Practice: Craft & Required Disclosure	& [ 0  1  3  6  8]\\
Art Practice: Music & Required Disclosure	& [ 0  1  2  6  7]\\
Art Practice: Digital & Required Disclosure	& [ 0  2  2  2  8]\\
Art Practice: Not Artist & Required Disclosure	& [ 0  1  1  5  3]\\
Art Practice: Other & Required Disclosure	& [ 0  0  3  2  4]\\
Art Practice: Tattoo & Required Disclosure	& [ 0  0  1  3  4]\\
Art Practice: Illustration & Required Disclosure	& [ 1  1  2  0  5]\\
Art Practice: Drawing &Required Disclosure	&  [ 0  0  1  4  2]\\
Art Practice: Sculpture &Required Disclosure &  [ 0  0  0  1  5]\\
Art Practice Maker & Required Disclosure	& [ 0  0  0  2  2]]\\
 
Art Practice: Paint &	AI Positive for Art	& [[23 26 33 35 44]\\
Art Practice: Photography & AI Positive for Art	&[ 8 14 29 35 23]\\
Art Practice: Writing & AI Positive for Art	&[12  7 12  7  8]\\
Art Practice: Designer & AI Positive for Art	&[ 5  7  7 11 10]\\
Art Practice: Craft & AI Positive for Art	&[ 2  4 12  0  0]\\
Art Practice: Music & AI Positive for Art	&[ 2  3  5  4  2]\\
Art Practice: Digital & AI Positive for Art	&[ 0  1 10  0  3]\\
Art Practice: Not Artist & AI Positive for Art	&[ 2  1  2  5  0]\\
Art Practice: Other & AI Positive for Art	&[ 1  1  4  3  0]\\
Art Practice: Tattoo & AI Positive for Art	&[ 0  0  3  0  5]\\
Art Practice: Illustration & AI Positive for Art &[ 2  1  2  2  2]\\
Art Practice: Drawing & AI Positive for Art	&[ 0  0  3  2  2]\\
Art Practice: Sculpture & AI Positive for Art	&[ 1  3  1  0  1]\\
Art Practice Maker & AI Positive for Art	&[ 2  1  1  0  0]]\\
 
Art Practice: Paint & AI Threat to Art & [[ 9 23 27 54 48]\\
Art Practice: Photography & AI Threat to Art &[ 5 17 22 35 30]\\
Art Practice: Writing & AI Threat to Art &[ 5  5  8 14 14]\\
Art Practice: Designer & AI Threat to Art &[ 4  4  6 15 11]\\
Art Practice: Craft & AI Threat to Art &[ 1  3  1  8  5]\\
Art Practice: Music & AI Threat to Art &[ 0  2  2  8  4]\\
Art Practice: Digital & AI Threat to Art &[ 1  1  6  5  1]\\
Art Practice: Not Artist & AI Threat to Art &[ 0  5  1  2  2]\\
Art Practice: Other & AI Threat to Art &[ 0  1  2  4  2]\\
Art Practice: Tattoo & AI Threat to Art &[ 0  1  1  3  3]\\
Art Practice: Illustration & AI Threat to Art &[ 2  1  0  2  4]\\
Art Practice: Drawing & AI Threat to Art &[ 1  1  1  3  1]\\
Art Practice: Sculpture & AI Threat to Art &[ 0  1  2  1  2]\\
Art Practice Maker & AI Threat to Art & [ 0  0  1  1  2]]\\
 \hline
\end{tabular}
\caption{Contingency table for the secondary variable of interest (Art Practice) and dependent variables on a 5-point Likert scale strongly disagree [(SD, leftmost number in bracket), disagree (D), neutral (N), agree (A), and strongly agree (SA, rightmost number in bracket)]. Outer brackets indicate the start and end of each contingency table, and the inner brackets indicate the start and end of a row in the table.}
\label{tab:CT_8}
\end{table*}

\begin{table*} 
\small
\centering
\tabcolsep=0.19cm
\begin{tabular}{@{}*{3}{c}@{}}
\textbf{Secondary Variable (Categorical)} & \textbf{Dependent Variable (Likert)} & \textbf{Contingency Table [SD D N A SA]} \\ \hline
Compensation: Tax  & Owner AI User	& [[ 0  2  2  2  0]\\
Compensation: None listed & Owner AI User	& [20  9  9  2  2]\\
Compensation: No profit needed but no profit for companies & Owner AI User	& [26 26 26 20  8]\\
Compensation: Flat fee & Owner AI User	& [ 5 13 14  8 14]\\
Compensation: Donate art to train models & Owner AI User	& [ 4  6 12 26 33]\\
Compensation: No profit needed but no profit for anyone & Owner AI User	& [ 5  9  3  7  3]\\
Compensation: Other & Owner AI User	& [ 3  0  4  1  2]\\
Compensation: Portion from users derivatives & Owner AI User	& [ 3 13  5  8  3]\\
Compensation: Portion of profit from anyone  & Owner AI User	& [ 4  7  5  9  7]\\
Compensation: Portion of profit from AI company & Owner AI User	& [16 14  9  7  5]\\
Compensation: Don't need profit & Owner AI User	& [ 1  2  6  6  7]]\\
 
Compensation: Tax  &	Owner Artist & [[ 0  1  1  3  1]\\
Compensation: None listed  &	Owner Artist &  [ 6 10 10  5 11]\\
Compensation: No profit needed but no profit for companies &	Owner Artist &  [11 32 29 23 11]\\
Compensation: Flat fee &	Owner Artist &  [ 5 16 13 11  9]\\
Compensation: Donate art to train models  &	Owner Artist &  [13 17  8 17 26]\\
Compensation: No profit needed but no profit for anyone &	Owner Artist &  [ 1  9  3  9  5]\\
Compensation: Other &	Owner Artist &  [ 1  3  4  1  1]\\
Compensation Portion from users derivatives &	Owner Artist &  [ 0  9 11  9  3]\\
Compensation: Portion of profit from anyone  &	Owner Artist &  [ 1  8 10  6  7]\\
Compensation: Portion of profit from AI company  &	Owner Artist &  [ 1  9 18 12 10]\\
Compensation &	Owner Artist &  [ 4  6  5  4  3]]\\
 
Compensation: Tax  &	Owner AI Company & [[ 0  3  1  2  0]\\
Compensation: None listed  &	Owner AI Company & [20 10  8  2  2]\\
Compensation: No profit needed but no profit for companies &	Owner AI Company & [39 33 18 10  6]\\
Compensation: Flat fee &	Owner AI Company & [10 12 12 13  7]\\
Compensation: Donate art to train models  &	Owner AI Company & [18  9 12 18 24]\\
Compensation: No profit needed but no profit for anyone &	Owner AI Company & [ 9 10  2  4  2]\\
Compensation: Other &	Owner AI Company & [ 6  1  3  0  0]\\
Compensation Portion from users derivatives &	Owner AI Company & [11 11  3  4  3]\\
Compensation: Portion of profit from anyone &	Owner AI Company & [ 6  8 10  5  3]\\
Compensation: Portion of profit from AI company  &	Owner AI Company & [23 12  6  7  3]\\
Compensation: Portion of profit from AI company  &	Owner AI Company & [ 4  4  7  4  3]]\\
\hline
\end{tabular}
\caption{Contingency table for the secondary variable of interest (Fair compensation) and dependent variables on a 5-point Likert scale strongly disagree [(SD, leftmost number in bracket), disagree (D), neutral (N), agree (A), and strongly agree (SA, rightmost number in bracket)]. Outer brackets indicate the start and end of each contingency table, and the inner brackets indicate the start and end of a row in the table.}
\label{tab:CT_10}
\end{table*}

\begin{table*} 
\small
\centering
\tabcolsep=0.19cm
\begin{tabular}{@{}*{3}{c}@{}}
\textbf{Secondary Variable (Categorical)} & \textbf{Dependent Variable (Likert)} & \textbf{Contingency Table [SD D N A SA]} \\ \hline
Compensation: Tax  &	Required Disclosure	& [[ 0  0  0  4  2]\\
Compensation: None listed  &	Required Disclosure	& [ 1  0  7  7 27]\\
Compensation: No profit needed but no profit for companies &	Required Disclosure	& [ 2  4 13 28 56]\\
Compensation: Flat fee &	Required Disclosure	& [ 2  4  2 26 19]\\
Compensation: Donate art to train models  &	Required Disclosure	& [ 3  8  8 27 35]\\
Compensation: No profit needed but no profit for anyone &	Required Disclosure	& [ 0  2  1 10 14]\\
Compensation: Other &	Required Disclosure	& [ 0  2  4  2  2]\\
Compensation Portion from users derivatives &	Required Disclosure	& [ 0  1  1 13 16]\\
Compensation: Portion of profit from anyone &	Required Disclosure	& [ 2  1  6  9 14]\\
Compensation: Portion of profit from AI company  &	Required Disclosure	& [ 0  0  5 17 29]\\
Compensation: Portion of profit from AI company &	Required Disclosure	& [ 3  2  8  3  6]]\\
 
Compensation: Tax  & AI Positive for Art & [[ 0  1  2  1  2]\\
Compensation: None listed  & AI Positive for Art & [17  8 12  2  3]\\
Compensation: No profit needed but no profit for companies & AI Positive for Art & [14 21 34 17 17]\\
Compensation: Flat fee & AI Positive for Art & [ 3  9 20 13  8]\\
Compensation: Donate art to train models  & AI Positive for Art & [ 1  1  7 29 42]\\
Compensation: No profit needed but no profit for anyone & AI Positive for Art & [ 6  2  6  9  4]\\
Compensation: Other & AI Positive for Art & [ 3  0  5  0  2]\\
Compensation Portion from users derivatives & AI Positive for Art & [ 3  8  8  9  3]\\
Compensation: Portion of profit from anyone & AI Positive for Art & [ 4  7  8  7  6]\\
Compensation: Portion of profit from AI company  & AI Positive for Art & [ 8 10 16 11  6]\\
Compensation: Portion of profit from AI company & AI Positive for Art & [ 1  2  6  6  7]]\\

Compensation: Tax  & AI Threat to Art	& [[ 0  1  4  1  0]\\
Compensation: None listed  & AI Threat to Art	& [ 2  2  7 11 20]\\
Compensation: No profit needed but no profit for companies & AI Threat to Art	& [ 3 16 15 43 26]\\
Compensation: Flat fee & AI Threat to Art	& [ 5  8  6 22 12]\\
Compensation: Donate art to train models  & AI Threat to Art	& [10 13 14 20 23]\\
Compensation: No profit needed but no profit for anyone & AI Threat to Art	& [ 2  5  4 12  4]\\
Compensation: Other & AI Threat to Art	& [ 0  2  5  2  1]\\
Compensation: Portion from users derivatives & AI Threat to Art	& [ 0  5  3 13 10]\\
Compensation: Portion of profit from anyone & AI Threat to Art	& [ 3  5  5  8 11]\\
Compensation: Portion of profit from AI company  & AI Threat to Art	& [ 0  7  9 18 17]\\
Compensation: Portion of profit from AI company & AI Threat to Art	& [ 3  1  8  5  5]]\\
\hline
\end{tabular}
\caption{Contingency table for the secondary variable of interest (Fair compensation) and dependent variables on a 5-point Likert scale strongly disagree [(SD, leftmost number in bracket), disagree (D), neutral (N), agree (A), and strongly agree (SA, rightmost number in bracket)]. Outer brackets indicate the start and end of each contingency table, and the inner brackets indicate the start and end of a row in the table.}
\label{tab:CT_11}
\end{table*}

\begin{table*} 
\small
\centering
\tabcolsep=0.19cm
\begin{tabular}{@{}*{3}{c}@{}}
\textbf{Secondary Variable (Categorical)} & \textbf{Dependent Variable (Likert)} & \textbf{Contingency Table [SD D N A SA]} \\ \hline
Gender Male & Owner AI User	 & [[36 54 61 62 60]\\
Gender Female & Owner AI User & [47 39 29 25 21]\\
Gender Non-binary & Owner AI User  & [ 4  5  4  7  3]]\\
 
Gender Male &	Owner Artist & [[23 66 63 63 57]\\
Gender Female & Owner Artist &[18 43 40 32 28]\\
Gender Non-binary & Owner Artist &[ 2  7  7  5  2]]\\
 
Gender Male &	Owner AI Company & [[71 59 60 43 40]\\
Gender Female & Owner AI Company &[65 41 20 23 12]\\
Gender Non-binary & Owner AI Company &[ 9  8  2  3  1]]\\
 
Gender Male &	Required Disclosure	& [[  6  15  35  95 122]\\
Gender Female & Required Disclosure	&[  6   7  17  45  86]\\
Gender Non-binary & Required Disclosure	& [  1   2   2   6  12]]\\
 
Gender Male &	AI Positive for Art	& [[21 41 80 68 62]\\
Gender Female & AI Positive for Art	& [35 26 38 29 33]\\
Gender Non-binary & AI Positive for Art	& [ 4  2  6  6  5]]\\
 
Gender Male &	AI Threat to Art & [[13 42 49 92 76]\\
Gender Female & AI Threat to Art &  [12 20 28 51 50]\\
Gender Non-binary & AI Threat to Art &  [ 3  3  2 12  3]]\\
\hline
\end{tabular}
\caption{Contingency table for the secondary variable of interest (Gender) and dependent variables on a 5-point Likert scale strongly disagree [(SD, leftmost number in bracket), disagree (D), neutral (N), agree (A), and strongly agree (SA, rightmost number in bracket)]. Outer brackets indicate the start and end of each contingency table, and the inner brackets indicate the start and end of a row in the table.}
\label{tab:CT_12}
\end{table*}

\begin{table*} 
\small
\centering
\tabcolsep=0.19cm
\begin{tabular}{@{}*{6}{c}@{}}
\textbf{} & \textbf{Strongly Disagree} & \textbf{Disagree} & \textbf{Neutral} & \textbf{Agree} & \textbf{Strongly Agree} \\ \hline
Strongly Disagree & 1	& 0	& 3 &	2 & 22\\
Disagree & 3	& 4	& 9	& 29 & 20\\\
Neutral &3	&6	&41&	17	&12\\
Agree& 11	&29	&59	&37&	18\\
Strongly Disagree& 42&	29&	12&	18	&27\\
\hline
\end{tabular}
\caption{Vertical side of table represents AI as a threat to the art workforce and the horizontal side of the table represents AI as a positive development for the field of art. Contingency table for the AI positive development for art and AI threat to art workforce on a 5-point Likert scale strongly disagree [(SD, leftmost number in bracket), disagree (D), neutral (N), agree (A), and strongly agree (SA, rightmost number in bracket)]. Outer brackets indicate the start and end of each contingency table, and the inner brackets indicate the start and end of a row in the table.}
\label{tab:CT_13}
\end{table*}

\begin{table*} 
\small
\centering
\tabcolsep=0.19cm
\begin{tabular}{@{}*{4}{c}@{}}
\textbf{Variable} & \textbf{p-value} & \textbf{H Statistic}  & \textbf{Degrees of Freedom} \\ \hline
Artist	&0.9873871802	&0.0002499081613&	1\\
Gender	&0.01738441229&	8.104362639&	2\\
Art Practice&	0.007501359367&	28.5814306&	13\\
AI Model Familiarity&	0.9798888491&	0.4306800685&	4\\
Age&	0.03938656365&	8.345451383&	3\\
Compensation&	1.05E-16&	98.58439653&	10\\
\hline
\end{tabular}
\caption{Results of the Kruskall-Wallis test for independence for the dependent variable (5-point Likert) AI art models are a positive development in the field of art. Strong evidence follows a threshold of a p-value less than 0.05.}
\label{tab:KW_positive}
\end{table*}

\begin{table*} 
\small
\centering
\tabcolsep=0.19cm
\begin{tabular}{@{}*{4}{c}@{}}
\textbf{Variable} & \textbf{p-value} & \textbf{H Statistic}  & \textbf{Degrees of Freedom} \\ \hline
Artist&	0.5445539432&	0.3671644759&	1\\
Gender&	0.5726378558&	1.115003553&	2\\
Art Practice&	0.9024842342&	6.993538631&	13\\
AI Model Familiarity&	0.4531083849&	3.665876796&	4\\
Age&	0.1005912351&	6.237926759&	3\\
Compensation&	0.04021463347&	19.003785&	10\\
\hline
\end{tabular}
\caption{Results of the Kruskall-Wallis test for independence for the dependent variable (5-point Likert) AI models are a threat to art workers. Strong evidence follows a threshold of a p-value less than 0.05.}
\label{tab:KW_threat}
\end{table*}

\begin{table*} 
\small
\centering
\tabcolsep=0.19cm
\begin{tabular}{@{}*{4}{c}@{}}
\textbf{Variable} & \textbf{p-value} & \textbf{H Statistic}  & \textbf{Degrees of Freedom} \\ \hline
Artist&	0.8660551636&	0.02845002072&	1\\
Gender&	0.326366829 &	2.239466576&	2\\
Art Practice&	0.8988829186&	7.062863624&	13\\
AI Model Familiarity&	7.67E-05&	24.08859461&	4\\
Age&	0.1962046718&	4.686999195&	3\\
Compensation&	0.001618196457&	28.29622161&	10\\
\hline
\end{tabular}
\caption{Results of the Kruskall-Wallis test for independence for the dependent variable (5-point Likert)  model creators should be required to disclose in detail what art \& images they use to train their AI models. Strong evidence follows a threshold of a p-value less than 0.05.}
\label{tab:KW_disclosure}
\end{table*}

\begin{table*} 
\small
\centering
\tabcolsep=0.19cm
\begin{tabular}{@{}*{4}{c}@{}}
\textbf{Variable} & \textbf{p-value} & \textbf{H Statistic}  & \textbf{Degrees of Freedom} \\ \hline
Artist&	0.795701369&	0.06703661173&	1\\
Gender&	5.41E-05&	19.65071632&	2\\
Art Practice&	0.1586078361&	17.97141505&	13\\
AI Model Familiarity&	0.1628188729&	6.531460606&	4\\
Age&	0.4668168207&	2.547225387&	3\\
Compensation&	5.36E-14&	84.94217137&	10\\
\end{tabular}
\caption{Results of the Kruskall-Wallis test for independence for the dependent variable (5 point Likert), if an AI art model was used by someone else to produce artwork recognizably in your style (e.g., in the style of Van Gogh) that work and its derivatives, should be considered the property of the person who used the AI model to generate the artwork. Strong evidence follows a threshold of a p-value less than 0.05.}
\label{tab:KW_user}
\end{table*}

\begin{table*} 
\small
\centering
\tabcolsep=0.19cm
\begin{tabular}{@{}*{4}{c}@{}}
\textbf{Variable} & \textbf{p-value} & \textbf{H Statistic}  & \textbf{Degrees of Freedom} \\ \hline
Artist&	0.2607376687&	1.264836076&	1\\
Gender&	0.2515206604&	2.760460289&	2\\
Art Practice&	0.05775284946&	21.84857601&	13\\
AI Model Familiarity&	0.2406398118&	5.489660662&	4\\
Age&	0.261387791&	4.000707054&	3\\
Compensation&	0.3765939846&	10.75907326&	10\\
\hline
\end{tabular}
\caption{Results of the Kruskall-Wallis  test for independence for the dependent variable (5 point Likert) If an AI art model was used by someone else to produce artwork recognizably in your style (e.g., in the style of Van Gogh and you are Van Gogh) that work and its derivatives should be considered the property of the artist. Strong evidence follows a threshold of a p-value less than 0.05.}
\label{tab:KW_artist}
\end{table*}

\begin{table*} 
\small
\centering
\tabcolsep=0.19cm
\begin{tabular}{@{}*{4}{c}@{}}
\textbf{Variable} & \textbf{p-value} & \textbf{H Statistic}  & \textbf{Degrees of Freedom} \\ \hline
Artist&	0.3626709363&	0.8286232383&	1\\
Gender&	0.0005455491796&	15.027435&2	2\\
Art Practice&	0.0004109645224&	37.0224258&	13\\
AI Model Familiarity&	4.31E-06&	30.27078455&	4\\
Age&	0.4650478321&	2.557169371&	3\\
Compensation&	1.34E-07&	51.62344111&	10\\
\hline
\end{tabular}
\caption{Results of the Kruskall-Wallis test for independence for the dependent variable (5 point Likert), if an AI art model was used by someone else to produce artwork recognizably in your style (e.g., in the style of Van Gogh) that work and its derivatives, should be considered the property of the AI model creators. Strong evidence follows a threshold of a p-value less than 0.05.}
\label{tab:KW_company}
\end{table*}

\begin{table*} 
\small
\centering
\tabcolsep=0.19cm
\begin{tabular}{@{}*{6}{c}@{}}
\textbf{Variable} & \textbf{p-value} & \textbf{Statistic}  & \textbf{Null mean}& \textbf{Null SD}& \textbf{Z-score} \\ \hline
Person of color	& 0.00206563&	243&	208.5620609&	11.17886095&	3.08063042\\
Professional Artist	&	0.08792146 &	472&	448.8618267&	13.55912215&	1.706465436\\
Purchased Art	&	0.82483924 &	855&	856.9180328&	8.666129548&	-0.2213251921\\
Used AI art	&	0.00000000&	603&	489.6674473&	13.59406494&	8.336914175\\
\hline
\end{tabular}
\caption{Results of the Cochran-Armitage test for independence for the dependent variable (5-point Likert) AI art models are a positive development in the field of art. Strong evidence follows a threshold of a p-value less than 0.05.}
\label{tab:CA_positive}
\end{table*}

\begin{table*} 
\small
\centering
\tabcolsep=0.19cm
\begin{tabular}{@{}*{6}{c}@{}}
\textbf{Variable} & \textbf{p-value} & \textbf{Statistic}  & \textbf{Null mean}& \textbf{Null SD}& \textbf{Z-score} \\ \hline
Person of color	&	0.27332660&	254&	242.6042155	&10.40298712&	1.09543388\\
Professional Artist	&	0.33653136&	510&	522.1264637&	12.61804523&	-0.9610413879\\
Purchased Art	&	0.47302764&	991&	996.7868852&	8.064652959&	-0.7175615956\\
Used AI art	&	0.00031337&	524&	569.5925059&	12.6505628&	-3.60399032\\
\hline
\end{tabular}
\caption{Results of the Cochran-Armitage test for independence for the dependent variable (5-point Likert) AI models are a threat to art workers. Strong evidence follows a threshold of a p-value less than 0.05.}
\label{tab:CA_threat}
\end{table*}

\begin{table*} 
\small
\centering
\tabcolsep=0.19cm
\begin{tabular}{@{}*{6}{c}@{}}
\textbf{Variable} & \textbf{p-value} & \textbf{Statistic}  & \textbf{Null mean}& \textbf{Null SD}& \textbf{Z-score} \\ \hline
Person of color	&	0.97821917&	293&	293.236534&	8.663735458	&-0.0273016136\\
Professional Artist &	0.08842447&	649&	631.0960187&	10.50846306&	1.703767826\\
Purchased Art &	0.22323397&	1213&	1204.819672&	6.716342048&	1.217973684\\
Used AI art	&	0.06462021&	669&	688.4683841&	10.53554409&	-1.847876475\\
\hline
\end{tabular}
\caption{Results of the Cochran-Armitage test for independence for the dependent variable (5-point Likert)  model creators should be required to disclose in detail what art \& images they use to train their AI models. Strong evidence follows a threshold of a p-value less than 0.05.}
\label{tab:CA_disclosure}
\end{table*}

\begin{table*} 
\small
\centering
\tabcolsep=0.19cm
\begin{tabular}{@{}*{6}{c}@{}}
\textbf{Variable} & \textbf{p-value} & \textbf{Statistic}  & \textbf{Null mean}& \textbf{Null SD}& \textbf{Z-score} \\ \hline
Person of color	&	0.00746442&	214&	182.2763466&	11.85761676	&2.675381911\\
Professional Artist	&	0.98389084&	392&	392.2903981&	14.38240218	&-0.02019121164\\
Purchased Art& 	0.15992987&	736&	748.9180328&	9.192317846	&-1.405307454\\
Used AI art&	0.00001229&	491&	427.9531616&	14.41946662	&4.372341923\\
\hline
\end{tabular}
\caption{Results of the Cochran-Armitage test for independence for the dependent variable (5 point Likert), if an AI art model was used by someone else to produce artwork recognizably in your style (e.g., in the style of Van Gogh) that work and its derivatives, should be considered the property of the person who used the AI model to generate the artwork. Strong evidence follows a threshold of a p-value less than 0.05.}
\label{tab:CA_user}
\end{table*}

\begin{table*} 
\small
\centering
\tabcolsep=0.19cm
\begin{tabular}{@{}*{6}{c}@{}}
\textbf{Variable} & \textbf{p-value} & \textbf{Statistic}  & \textbf{Null mean}& \textbf{Null SD}& \textbf{Z-score} \\ \hline
Person of color	& 	0.00037047 &	237&	198.4355972&	10.83186977&	3.560272014\\
Professional Artist &	0.59775878&	434&	427.0679157&	13.13824781&	0.5276262413\\
Purchased Art	&	0.74883816&	818&	815.3114754&	8.397133403&	0.3201717135\\
Used AI art	&	0.08222217&	443&	465.8922717&	13.17210597&	-1.737935582\\
\hline
\end{tabular}
\caption{Results of the Cochran-Armitage test for independence for the dependent variable (5 point Likert) If an AI art model was used by someone else to produce artwork recognizably in your style (e.g., in the style of Van Gogh and you are Van Gogh) that work and its derivatives should be considered the property of the artist. Strong evidence follows a threshold of a p-value less than 0.05.}
\label{tab:CA_artist}
\end{table*}

\begin{table*} 
\small
\centering
\tabcolsep=0.19cm
\begin{tabular}{@{}*{6}{c}@{}}
\textbf{Variable} & \textbf{p-value} & \textbf{Statistic}  & \textbf{Null mean}& \textbf{Null SD}& \textbf{Z-score} \\ \hline
Person of color	&	0.00001507&	188&	137.030445&	11.77770463	&4.327630607\\
Professional Artist	&	0.02548492&	263&	294.9133489&	14.28547474	&-2.233971885\\
Purchased Art&	0.18814356&	551&	563.0163934&	9.130367983	&-1.316090815\\
Used AI art	&	0.68942716&	316&	321.7236534&	14.32228939	&-0.3996325755\\
\hline
\end{tabular}
\caption{Results of the Cochran-Armitage test for independence for the dependent variable (5 point Likert), if an AI art model was used by someone else to produce artwork recognizably in your style (e.g., in the style of Van Gogh) that work and its derivatives, should be considered the property of the AI model creators. Strong evidence follows a threshold of a p-value less than 0.05.}
\label{tab:CA_company}
\end{table*}

\section*{Appendix F: Code and Data Availability}

All data and all code are available from our GitHub repository, which can be found at the following URL: 

\texttt{https://github.com/juniperlovato/AASP}

% %%%%%%%%%%%%%%%%%%%%

\end{document}